\documentclass[nonacm,sigplan]{acmart}

\usepackage{siunitx}
\usepackage{tikz}
\usepackage{amsmath}

\usepackage{graphicx}
\usepackage{float} 
\usepackage{subcaption}

\usepackage{enumerate}
\usepackage[ruled,vlined,linesnumbered]{algorithm2e}

\usepackage{changepage} % for adjusting margins

\usepackage{listings}

\usepackage{amssymb}
\usepackage{gensymb}
\usepackage{booktabs}

\usepackage{makecell}
\usepackage{colortbl}

\usepackage{multirow}

\usepackage{soul}
\soulregister{\cite}7
\soulregister{\citep}7
\soulregister{\citet}7
\soulregister{\ref}7
\soulregister{\pageref}7

\usepackage{color, xcolor}

\definecolor{codegreen}{rgb}{0,0.6,0}
\definecolor{codegray}{rgb}{0.5,0.5,0.5}
\definecolor{codepurple}{rgb}{0.58,0,0.82}
\definecolor{backcolour}{rgb}{0.95,0.95,0.92}

\lstdefinestyle{mystyle}{
    commentstyle=\color{codegreen},
    keywordstyle=\color{magenta},
    numberstyle=\tiny\color{codegray},
    stringstyle=\color{codepurple},
    basicstyle=\footnotesize,
    breakatwhitespace=false,         
    breaklines=true,                 
    captionpos=b,                    
    keepspaces=true,                 
    numbers=left,                    
    numbersep=5pt,                  
    showspaces=false,                
    showstringspaces=false,
    showtabs=false,                  
    tabsize=2
}

\begin{document}

\title{\fontsize{16}{20}\selectfont \bf LOOPRAG: Enhancing Loop Transformation Optimization with Retrieval-Augmented Large Language Models}

\author{Yijie Zhi}
\affiliation{%
  \institution{Zhejiang University}
  \city{Hangzhou}
  \state{Zhejiang}
  \country{China}
}

\author{Yayu Cao}
\affiliation{%
  \institution{Zhejiang University}
  \city{Hangzhou}
  \state{Zhejiang}
  \country{China}
}

\author{Jianhua Dai}
\affiliation{%
  \institution{Zhejiang Institute of Administration}
  \city{Hangzhou}
  \state{Zhejiang}
  \country{China}
}

\author{Xiaoyang Han}
\affiliation{%
  \institution{Zhejiang University}
  \city{Hangzhou}
  \state{Zhejiang}
  \country{China}
}

\author{Jingwen Pu}
\affiliation{%
  \institution{Zhejiang University}
  \city{Hangzhou}
  \state{Zhejiang}
  \country{China}
}

\author{Qingran Wu}
\affiliation{%
  \institution{Zhejiang University}
  \city{Hangzhou}
  \state{Zhejiang}
  \country{China}
}

\author{Sheng Cheng}
\affiliation{%
  \institution{Beijing ShenZhou Aerospace Software Technology Ltd.}
  \city{Beijing}
  \state{Beijing}
  \country{China}
}

\author{Ming Cai}
% \authornote{Corresponding author}  % 标记通讯作者
% \email{mingcai@zju.edu.cn}  % 通讯作者邮箱放在这里
\affiliation{%
  \institution{Zhejiang University}
  \city{Hangzhou}
  \state{Zhejiang}
  \country{China}
}

% 如果需要简短的作者列表用于页眉
\renewcommand{\shortauthors}{Zhi et al.}

\begin{abstract}
% 直接说是reasoning task
Loop transformations are semantics-preserving optimization techniques applied at the source level, widely used in compilers to maximize objectives such as vectorization and parallelism. Despite decades of research, applying the optimal composition of loop transformations remains challenging due to inherent complexities, including dependency analysis and cost modeling for optimization objectives.

Recent studies have explored the potential of Large Language Models (LLMs) for code optimization. However, our key observation is that LLMs often struggle with effective loop transformation optimization, frequently leading to errors or suboptimal optimization, thereby missing significant opportunities for performance improvements.

% To bridge the gap, we propose LOOPRAG, a novel retrieval-augmented generation framework that guides LLMs in performing loop optimization optimization. In LOOPRAG, we introduce a parameter-driven method to generate diverse example codes and optimize them through loop transformations. To retrieve the most informative example as demonstrations for LLMs, we present a loop-aware algorithm that balance similarity and diversity for retrieval. To ensure correct and efficient code generation, we utilize a feedback-based iterative mechanism that incorporates results from compilation, execution and evaluation.

\sloppy To bridge this gap, we propose LOOPRAG, a novel retrieval-augmented generation framework designed to guide LLMs in performing effective loop optimization on Static Control Part (SCoP). We introduce a parameter-driven method to harness loop properties, which trigger various loop transformations, and generate diverse yet legal example codes serving as a demonstration source. 
To effectively obtain the most informative demonstrations, we propose a loop-aware algorithm based on loop features, which balances similarity and diversity for code retrieval. To enhance correct and efficient code generation, we introduce a feedback-based iterative mechanism that incorporates compilation, testing and performance results as feedback to guide LLMs. Each optimized code generated by LOOPRAG undergoes mutation, coverage and differential testing to perform equivalence checking.

We evaluate LOOPRAG on PolyBench, TSVC and LORE benchmark suites, and compare it against compilers (GCC-Graphite, Clang-Polly, Perspective and ICX) and representative LLMs (DeepSeek and GPT-4). The results demonstrate average speedups over compilers of up to 11.20$\times$, 14.34$\times$, and 9.29$\times$ for PolyBench, TSVC, and LORE, respectively, and speedups over base LLMs of up to 11.97$\times$, 5.61$\times$, and 11.59$\times$.

\end{abstract}

\maketitle % should come after the abstract
\pagestyle{plain} % should come right after \maketitle

\section{Introduction} 
Loop transformation is a critical task in code optimization research, altering the execution order of statements in a semantics-preserving manner at source level to improve performance. However, applying the optimal composition of loop transformations remains a long-standing challenge, that requires a deeper understanding of the code's logic and greater context awareness \cite{10606318, cummins2023largelanguagemodelscompiler}.

%Code optimization has long been concerned in compilation research. Loop transformations can improve the performance in semantics-preserving way by changing the execution order of the statements, making it a significant technique. 
%However, finding the optimal compositions of loop transformations has long been a problem since compilers may not capture these optimizations which require a deeper understanding of the code's logic and context \cite{10606318, cummins2023largelanguagemodelscompiler}.
% Past research developed polyhedral model (i.e., polytope model) \cite{feautrier2005automatic, lengauer1993loop} to formulate this problem by representing loop iterations as points in a geometric space and searching the exploration of various transformation strategies to achieve multiple objectives, such as maximizing parallelism and cache efficiency. However, different optimization objectives are partly contradictory, thus needs to design complex cost models or heuristics in solving this linear programming problem.

% 第三段insight是在现有生成能力上加强分析能力（借助传统编译器的能力）
% 第四段however，细节问题

With the advancement of Large Language Models (LLMs), recent studies have explored their high-level semantic understanding and contextual analysis capabilities for code optimization, such as auto-vectorization \cite{taneja2024llmvectorizerllmbasedverifiedloop} and OpenMP directives prediction \cite{rosas2024aioptimizecodecomparative}. Their findings indicate that LLMs demonstrate potential in code optimization tasks. However, their current capabilities are not yet sufficient, and significant improvements in correctness and performance are required.

\begin{figure}[tb]
	\centering
	\includegraphics[width=0.48\textwidth]{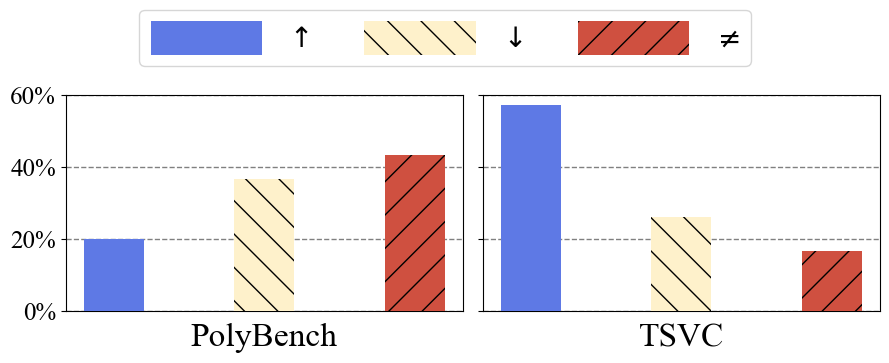}
	\caption{Percentage of code optimization results generated by GPT-4 compared to PLuTo on PolyBench \cite{pouchet2023PolyBench} and TSVC \cite{44642}. All codes are compiled using GCC. $\uparrow$ and $\downarrow$ indicate performance that outperforms and underperforms PLuTo, respectively. $\neq$ represents code that is not equivalent in semantics to the original.}
	\label{fig:1}%文中引用该图片代号
\end{figure}

% Motivated by the power of In-Context Learning (ICL) \cite{NEURIPS2020_1457c0d6}, we experiment to use GPT-4 \cite{openai2024gpt4technicalreport} to achieve loop transformation optimization for codes in C language (Figure \ref{fig:1}).  
% Our key finding is that LLMs are capable of generating loop codes with correct syntax, but they frequently miss optimization opportunities for conducting loop transformations, which results in suboptimal optimization or is not equivalent in semantics. This aligns with the findings in Tehrani et.al. \cite{10.1145/3613905.3650896} and Rosas et.al. \cite{rosas2024aioptimizecodecomparative}.
% A possible explanation is that LLMs lack intrinsic mechanisms (e.g., cost models or heuristics found in compilers) to estimate and predict potential performance gains, which limits the ability of applying profitable loop transformations. 

% Meanwhile, adding demonstrations in prompts, which suggests appropriate composition of loop transformations for target code, can guide LLMs to generate optimized code with higher efficiency. For example, Listing \ref{lst:1} in §\ref{2.2} can achieve 71.60$\times$ speedup as compared to the result generated by GPT-4 through demonstrations in §\ref{6.5}.
% Hence, we need example codes and their corresponding optimization results to construct demonstrations, which help LLMs maximize the ability of applying profitable loop transformation optimization.

To evaluate the performance of LLMs compared to compilers, we conduct an empirical study by prompting GPT-4 \cite{openai2024gpt4technicalreport} and employing the source-to-source optimization compiler PLuTo \cite{10.1145/1379022.1375595} to perform loop transformation on Static Control Part (SCoP) in C program (Figure \ref{fig:1}). The results indicate that, compared to PLuTo, GPT-4 frequently misses optimization opportunities for achieving optimal loop transformations and struggles to generate semantically equivalent programs. A possible explanation is that LLMs lack intrinsic mechanisms (e.g., cost models or heuristics in compilers) to predict performance impact, which limits their abilities of applying profitable loop transformations. 

Motivated by In-Context Learning (ICL) \cite{NEURIPS2020_1457c0d6}, the optimization ability of LLMs can be enhanced by incorporating informative demonstrations into the prompt, guiding them to learn potential optimal transformation sequences for the target code. For example, Listing \ref{lst:1} in §\ref{2.2} can achieve 71.60$\times$ speedup as compared to the result generated by GPT-4 in Figure \ref{fig:1} through demonstrations in §\ref{6.5}. Hence, we highlight the critical role of providing informative demonstrations to guide LLMs in performing efficient code optimization.

% existing不够various -> loop properties -> contradiction -> parameter-driven

% Generally, loop transformations encompass a wide variety of types that can be combined in different ways. 
% The vast search space for finding the optimal composition of loop transformations highlights the need of various examples. 
As a result, a large-scale dataset containing diverse example codes and their optimized versions in SCoP generated by an optimization compiler using various loop transformations is an essential resource and can serve as a demonstration source.
Although there are plenty of loop code datasets \cite{8167779, 9370322} and loop generators \cite{10.1145/3591295, berezov_et_al:OASIcs.PARMA-DITAM.2022.3}, they are limited in providing compositions of loop transformations, resulting in a lack of diverse demonstrations. 
% For examples, the code produced by COLA-Gen \cite{berezov_et_al:OASIcs.PARMA-DITAM.2022.3} can only support 3 out of 6 types of loop transformations. 
Based on the observation that diverse loop properties (e.g., loop structure and data dependence) can trigger various loop transformations, we aim to produce example codes at scale by incorporating loop properties with diversity. However, due to interdependencies among loop properties, contradictions arise when the values of loop properties are assigned randomly.
To solve this problem, we propose a \textbf{parameter-driven method} to systematically harness loop properties through loop parameters and generate diverse yet legal example codes.
% a comprehensive dataset, thereby providing effective examples for LLMs and enhancing the diversity of demonstrations. 

% Only text retrieval -> loop-properties -> redundant -> loop-aware + feature

% However, it is non-trivial to retrieve the most appropriate example codes
Another significant challenge is to identify the most appropriate example codes in SCoP. 
We aim to retrieve example codes based on the semantic similarity between example and target code. However, existing retrieval methods (e.g., BM25 \cite{INR-019}, GIST \cite{solatorio2024gistembedguidedinsampleselection} and BGE \cite{xiao2024cpackpackedresourcesgeneral}) prioritize text features over the code semantics. 
Loop properties can reflect semantic information relevant to loop transformation, but using all of them in retrieval may lead to redundancy.
Therefore, we extract loop features from loop properties and present a \textbf{loop-aware algorithm} that balances similarity and diversity for retrieval. 
These retrieved example codes and their optimized versions are integrated into demonstrations to prompt LLMs in generating optimized codes.

% In order to correctness and performance->3个feedback->testing not real-world->compiler testing for equivalence checking

To enhance correct and efficient code generation, we introduce a \textbf{feedback-based iterative mechanism} that incorporates compilation results, testing results and performance rankings into prompts as informative feedback to guide LLMs.
% for loop transformation optimization in LLMs and iteratively improve the correctness and performance of the final generated code.
These three types of feedback are obtained during iterative generation process. 
% through compilation, execution and evaluation testing. 
However, verifying whether loop transformations preserve the original semantics is challenging, as proving semantic equivalence is a long-standing and undecidable problem \cite{dramko2025fastfinegrainedequivalencechecking}. Existing methods \cite{10.1145/3453483.3454030, 10.1145/3624062.3624140} are still unable to handle real-world programs with complex structures or address the black-box nature of LLM generation. Therefore, instead of relying on theoretical proofs that account for data and control dependencies, we directly test the specific generated outputs while performing optimization through LLMs. In order to cover as many corner cases as possible, we adopt techniques from compiler testing, including mutation, coverage and differential testing for semantic equivalence checking.

% We firstly construct a dataset of example codes. The formatted example codes are created considering loop properties, which makes it potential in generating diverse optimized version of code with profitable compositions of loop transformations. The optimized version of codes and the information of above properties could be obtained through existing optimizing compilers and analyzers. These data provide sufficient information for selecting appropriate demonstration for rag usage.

% Thus, we design our loop-aware algorithm to match these key information between example codes and target code. 
% Then, the composition of loop transformations used in optimized version of retrieved example codes will be added into prompt. 
% These loop transformations demonstrations could significantly exploit the ability of LLMs to analyze, learn and use appropriate loop transformations to generate optimized code and improve its performance.

% The compilation results, testing results and performance rankings of the intermediate generated codes will be successively returned to prompt, which can guide LLMs learn from those code with errors and lower performance, and prevent the optimization strategy in regeneration. At last, we also design an exhaustive procedure for differential, mutation and coverage testing by randomly initialization \cite{NEURIPS2023_43e9d647}, and use the output of computed array as test oracle to defect the correctness of semantics equivalence.

\sloppy In this paper, we propose LOOPRAG, a novel retrieval-augmented generation (RAG) framework for loop transformation optimization in SCoP. 
LOOPRAG aims to guide LLMs in learning potential optimal transformations through demonstrations while preserving their inherent optimization capabilities.
We evaluate LOOPRAG by comparing it with four compilers (GCC-Graphite\cite{pop2006graphite}, Clang-Polly\cite{grosser2012polly}, Perspective\cite{apostolakis2020perspective} and ICX), two representative LLMs (DeepSeek \cite{deepseekai2024deepseekv2strongeconomicalefficient} and GPT-4), and two LLM-based models (PCAOT \cite{rosas2024aioptimizecodecomparative} and LLM-Vectorizer \cite{taneja2024llmvectorizerllmbasedverifiedloop}), respectively. Our approach achieves a significant boost in program execution efficiency. For compilers, the average speedups reach up to $11.20\times$, $14.34\times$, and $9.29\times$ on PolyBench, TSVC, and LORE \cite{8167779}, respectively. For base LLMs, the average speedups are up to $11.97\times$, $5.61\times$, and $11.59\times$ on the same benchmarks. For LLM-based models, the average speedups are up to $8.10\times$ and $5.97\times$ on PolyBench and TSVC, respectively.
% \vspace{0.5cm} % 调整第一章后的间距

% 第二章前后调整间距
% \vspace{-0.75cm} % 调整第二章前的间距
\section{Background and Motivation} \label{2}

% loop transformation种类，以及和上述特征的关联性，辅以详细例子说明直接为什么不行，以及loop transformation的提示下有什么变化从而进行优化（对应challenge 1）

% 循环多样性（语句，层级，结构），依赖多样性（指向，方向），数组多样性（下标），以及和loop transformation有何联系，从而说明构造数据要求和检索关注对象（对应challenge2）

\subsection{Loop Properties and SCoP} \label{2.1}
The modeling of loops is essential when synthesizing diverse example codes. There are three types of loop properties worth concerning in the modeling process.

The first type refers to loop structure, which includes the number of statements, loop bounds, loop depth (i.e., loop levels), and loop schedule. Loop schedule \cite{10.5555/556139} defines the execution order of statements and can be represented in various forms (e.g., 2d+1 form and schedule tree \cite{verdoolaege:hal-00911894}). The value in constant dimension of loop schedule decides whether the loop is \emph{perfect}, where all statements are located in the innermost loop, or \emph{imperfect}, where not all statements reside in the innermost level.

The second type is related to data dependence, considering the number of dependence, dependence type and dependence distance \cite{1130282272927219456, 10.1007/BFb0025880}. Based on the relationship between operations, there are three types of data dependence: RAW (read after write), WAW (write after write), and WAR (write after read). Regarding the relationship between iterations, the dependence distance characterizes the difference between the iteration vectors of two dependent operations. 
This difference also categorizes dependence into two types: loop-independent dependence, which exists within the same iteration of a loop, and loop-carried dependence, which occurs across different iterations \cite{wolfe1978techniques}.

The third type focuses on array access, which includes the number of arrays, array name, array size and indexes (i.e., access functions). Array indexes play a crucial role in determining data locality \cite{742790, 10.1145/3178372.3179507}, as the access order of arrays directly impacts the efficiency of memory and cache usage.

These three types of loop properties can be leveraged to model loops in Static Control Part (SCoP) \cite{10.1007/978-3-540-24644-2_14}. A SCoP represents a well-structured segment of a program in which all loop bounds and conditional statements are defined by affine functions of surrounding iterators and global parameters. Within SCoP, only \emph{for} loops and \emph{if} conditionals are permitted. A defining characteristic of a SCoP is that it is side-effect free, meaning that within the SCoP, all operations (including function calls) are restricted to expressions that do not modify any state outside the SCoP. Owing to this mathematical tractability, SCoPs can be analyzed and optimized using static code analyzers (e.g., Clan \cite{10.1007/978-3-540-24644-2_14}, ISL \cite{Verdoolaege2010isl}, and pet \cite{17bd1f6549cc4aca85b60c3e72ef04be}) as well as optimization compilers (e.g., PLuTo \cite{10.1145/1379022.1375595} and PPCG \cite{10.1145/2400682.2400713}).

\begin{figure}[tb]
  \centering
  \begin{minipage}{0.48\textwidth}
\hspace{-1.2em}\includegraphics[width=1.03\textwidth]{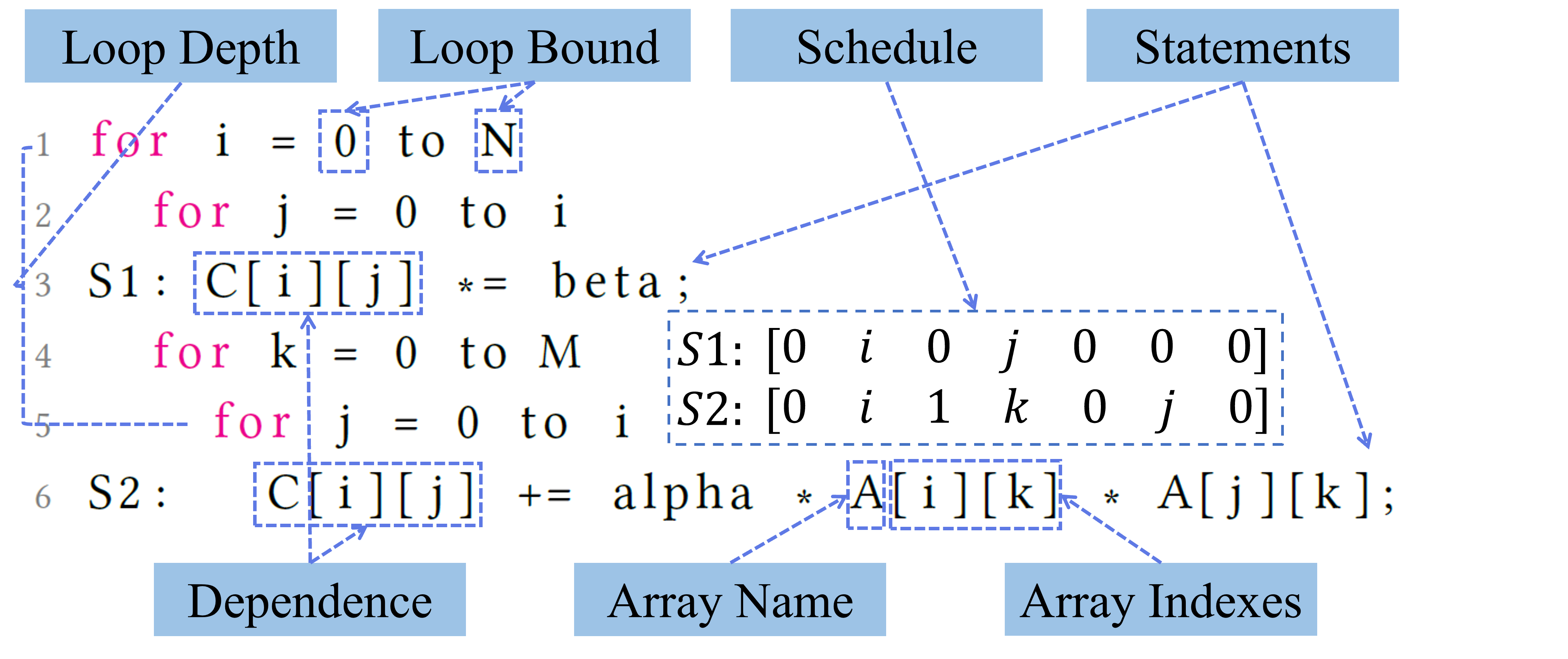}
    \captionsetup{labelformat=default}
	\caption{Original code \emph{syrk}.}
	\label{fig:2}
% \begin{lstlisting}[language=C, caption=Original Code \emph{syrk}, label=lst:0.1]
% for i = 0 to N
%   for j = 0 to i
% S1: C[i][j] *= beta;
%   for k = 0 to M
%     for j = 0 to i   
% S2:   C[i][j] += alpha * A[i][k] * A[j][k];
% \end{lstlisting}
  \end{minipage}
  \hfill
  \begin{minipage}{0.48\textwidth}
\begin{lstlisting}[language=C, caption=Optimized code \emph{syrk}., label=lst:1, captionpos=t]
#pragma omp parallel
for t1 = 0 to N/32
 for t2 = 0 to t1
  for t3 = 32*t1 to 32*t1+32
   for t4 = 32*t2 to 32*t2+32
S1: C[t3][t4] *= beta;
    for k = 0 to M
S2:  C[t3][t4] += alpha * A[t3][k] * A[t4][k];
\end{lstlisting}
  \end{minipage}
\end{figure}

Figure \ref{fig:2} illustrates a SCoP example derived from \emph{syrk} in PolyBench. The SCoP contains two statements, $S1$ and $S2$, with loop depth in $2$ and $3$, respectively. The loop bounds for iterator $i$, $j$ and $k$ are $0$ to $N$, $0$ to $i$ and $0$ to $M$, which constrain the legal iteration space. The schedules are presented in 2d+1 form with $S1:[0,i,0,j,0,0,0]$ and $S2:[0,i,1,k,0,j,0]$, which indicates that $S1$ and $S2$ share the same loop $i$ but $S1$ is executed before $S2$ within loop $i$. Two arrays, $C$ and $A$, are referenced in this SCoP. The array indexes are $[i,j]$ for array $C$ and $[i,k]$, $[j,k]$ for array $A$. The operations $*=$ and $+=$ induce three types of dependencies on array $C$: WAW, WAR and RAW.

These properties serve as the foundational information of defining a loop, and can trigger various loop transformations.

% enabling efficient and large-scale generation of diverse example codes. 

\subsection{Loop Transformations} \label{2.2}
Loop transformations refer to a set of techniques (e.g., loop interchange, loop tiling, loop fusion, loop distribution, loop skewing, loop shifting) for performance optimization (e.g., parallelism, vectorization, data locality) \cite{girbal2006semi}. These optimizations must carefully analyze the loop properties: loop structure, data dependence and array access.
Listing \ref{lst:1} illustrates an optimized code $syrk$ generated by GPT-4 for original code in Figure \ref{fig:2} when prompted with informative demonstrations.

This optimized code utilizes three types of loop transformations: loop tiling, loop fusion, and loop interchange, which are absent in the code generated by GPT-4 without demonstrations.
% 一上来直接llm原本失败了，而我们的优化了；接下来我们做成了哪些变换导致了优化；接下来我们的demonstration提供了哪些使得我们能成而llm不能；最后，刺激我们想找到最匹配的demonstration
Loop tiling divides the iterator $i$ by a block length of $32$, and separates it into iterations in loop $t1$ and $t3$, which make array access in the block more efficient for data locality in cache usage. The iterator $j$, $t2$ and $t4$ follow a similar approach.

In this optimized code, loop interchange transforms the schedule of $S2$ by exchanging its surrounding iterators $k$ and $j$. $S2$ is located within loop $k$ instead of $j$ (i.e., $t2$ and $t4$). Loop fusion then groups $S1$ and $S2$ into the same loop $t4$. 
These two loop transformations combine to enhance self-spatial and self-temporal data locality \cite{10.1145/989393.989437}. 
Furthermore, the $pramga$ for OpenMP \cite{660313} directive instructs the compiler to achieve parallelism for the SCoP.

These optimizations in the code generated above are attributed to appropriate demonstrations delivered to GPT-4, achieving 71.60$\times$ speedup compared to GPT-4 without demonstrations.

\section{Overview}

\begin{figure}[tb]
    \centering
    \includegraphics[width=\linewidth]{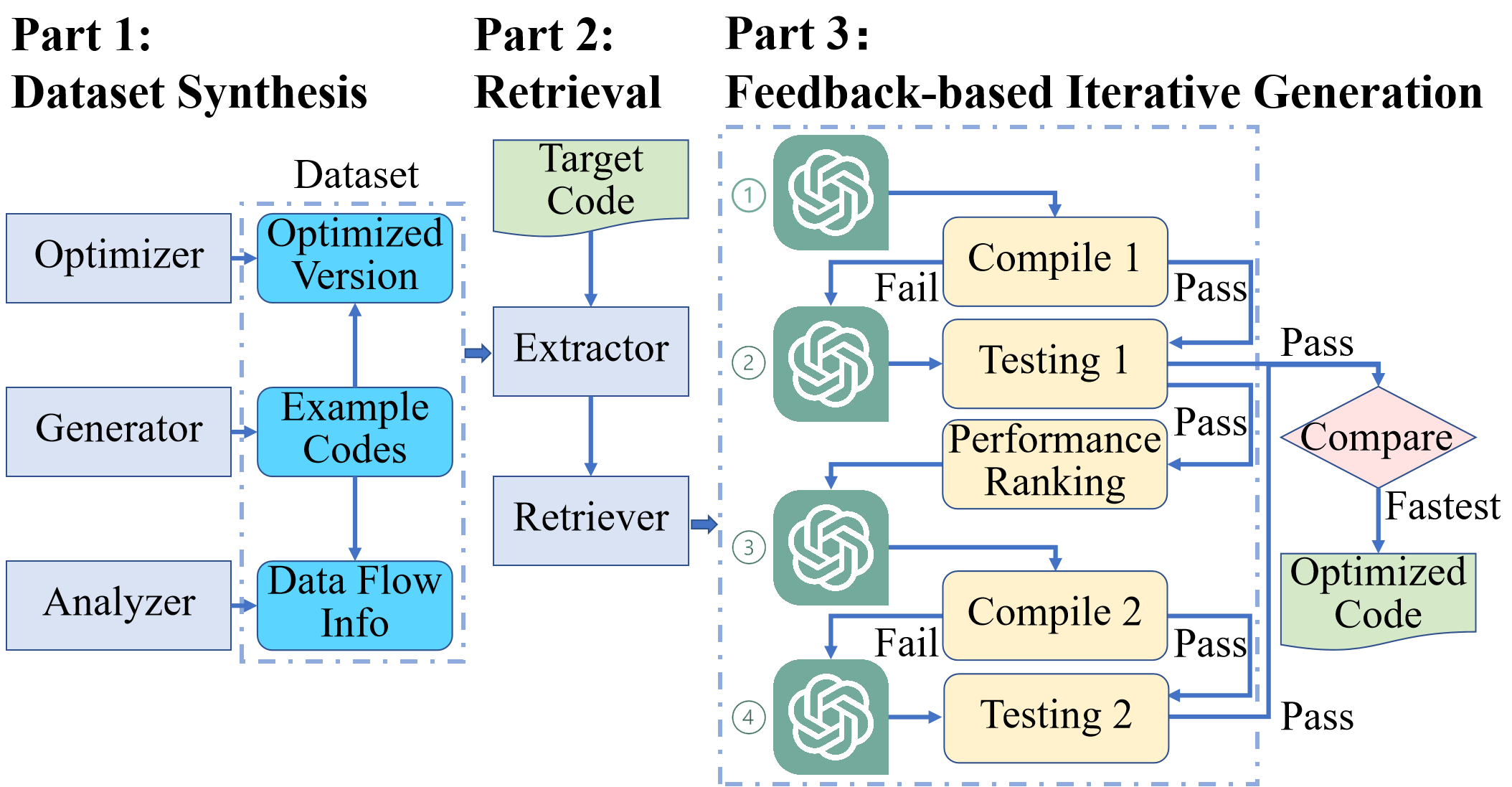}
    \caption{LOOPRAG overview.}
    \label{fig:3}
\end{figure}

% 要说清楚我们的创新步骤的目的，例如analysis的目的forretrieval，rerank for diverse and accurate，feedback for tradeoff between correctness and performance
LOOPRAG is a source-to-source loop transformation optimization framework for C language. 
Figure \ref{fig:3} illustrates the overview of LOOPRAG, encompassing three parts: \textbf{Dataset Synthesis}, \textbf{Retrieval}, and \textbf{Feedback-based Iterative Generation}. 

Initially, LOOPRAG introduces a parameter-driven method to configure loop properties and synthesize example codes. These examples are optimized through a source-to-source optimization compiler, and the data flow information is extracted from the example codes using analyzers. The example codes, along with their corresponding optimized versions and data flow information, are subsequently stored as the synthesized dataset (§\ref{4.1}).

Next, LOOPRAG extracts SCoPs from example codes in synthesized dataset and target code, and employs a loop-aware algorithm to retrieve appropriate example SCoPs for target SCoP. This algorithm leverages loop features extracted from loop properties to identify both syntactic and semantic similarity between example and target. The top-N retrieved results are then selected with their optimized versions of SCoPs as demonstrations to prompt LLMs (§\ref{4.2}).

To improve the correctness and performance, LOOPRAG applies feedback-based iterative mechanism in code generation. The whole procedure consists of four iterations, utilizing feedback from compilation, testing and performance results. During generation, each generated SCoP is integrated back into the program and then checked for equivalence using mutation, coverage and differential testing. Finally, the passing codes are compared for performance, and the fastest one is output (§\ref{4.3}).

\section{Approach}

\subsection{Dataset Synthesis} \label{4.1}
%0.75
% 开头直接说明现有的数据集不够复杂是为什么，用scop，dependence等信息说明，所以必须构造，

There are several existing loop code datasets (e.g., ANGHABENCH \cite{9370322}) and loop generators (e.g., COLA-Gen \cite{berezov_et_al:OASIcs.PARMA-DITAM.2022.3}) in C language. However, none of them can provide example codes with various loop properties, limiting the ability of LLMs to learn varied compositions of loop transformations. For example, COLA-Gen generates codes by mutating loop depth and number of arrays, but it only supports a single statement in \emph{perfect} loop with loop-carried dependence. Hence, it cannot apply loop fusion, loop distribution or loop shifting, thereby failing to demonstrate complex compositions of loop transformations (§\ref{6.3.1}).

% 引入dataset的构造流程（generator+compiler+analyzer）
Based on the observation that diverse loop properties can trigger various loop transformations, we aim to produce example codes at scale by incorporating diverse loop properties. However, randomly assigning each loop property can lead to contradictions due to interdependencies among them. For example, if a WAW dependence on array $A$ is established for a statement, it necessitates a corresponding write to an array. However, if array access is set independently, it may violate this constraint to acquire multiple writes to array, which could also result from other WAR or RAW dependence. Similarly, setting loop bounds and array sizes separately may cause an "array index out of bounds" error if the loop's upper bound exceeds the array size.

\begin{figure}[tb]
    \centering
    \includegraphics[width=\linewidth]{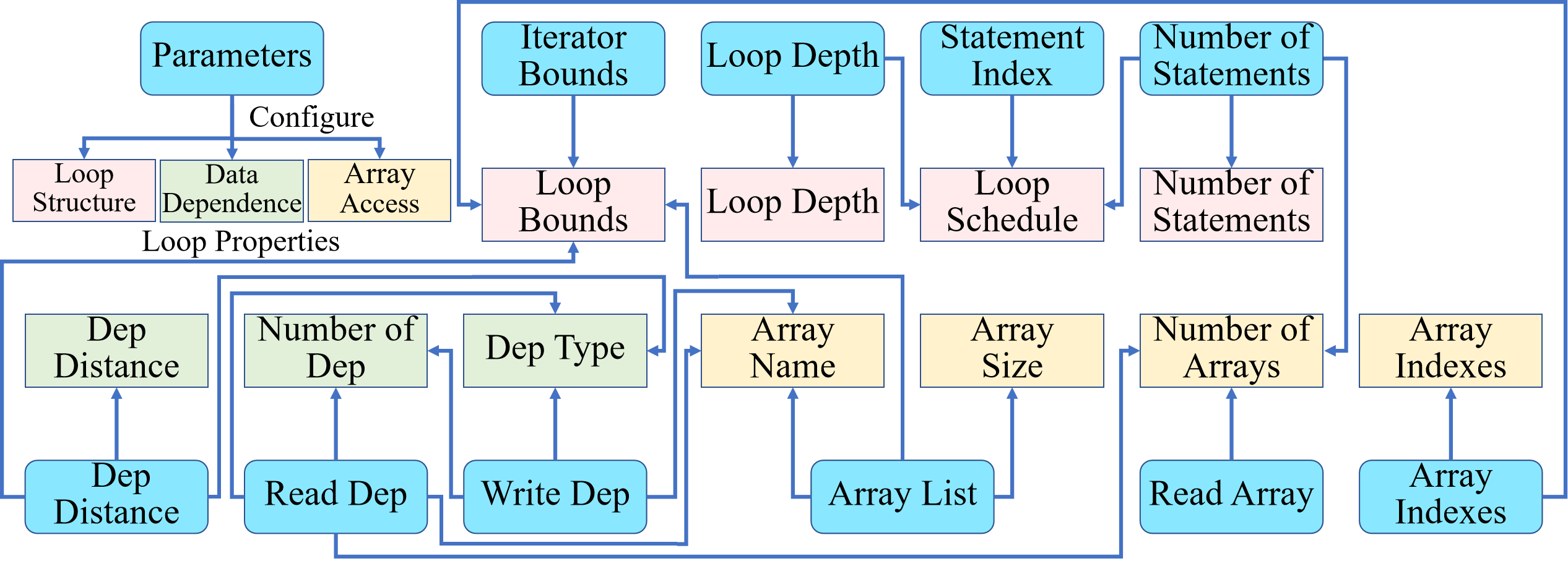}
    \caption{The relationship between parameters and loop properties. The directed line segment from a parameter to a loop property indicates that the parameter can control the loop property. A loop property may be configured by one or more parameters.}
    \label{fig:4.1}
\end{figure}

% a. 对于冲突：
% 	i. 解耦拆分（e.g. bounds）
% 	ii. 不同properties设置优先级，相同properties再check（是否成环检测，如果冲突根据schedule顺序保留前者）(e.g., )
% b. 对于独立单独处理

To address this problem, we introduce a parameter-driven method that uses loop parameters to configure loop properties, resolving contradictions by decoupling, priority-based assignment and contradiction-check mechanisms. We assign the task of constructing eleven loop properties to ten parameters, and create the SCoP for synthesized codes. The relationship between parameters and loop properties is shown in Figure \ref{fig:4.1}. More details are discussed in Appendix \ref{10.1}.

We begin by separating interdependent loop properties into distinct groups. For each group, we identify influencing factors among the correlated properties. Then, we abstract the influencing factors as fine-grained parameters to decouple their relationships and constrain the correlated properties. For example, loop bounds can be defined by four parameters: \textbf{Iterator  Bound}, \textbf{Array List}, \textbf{ Dep Distance} and \textbf{Array Indexes}. If we select one array $C[M][N]$ (where $M<N-1$) from \textbf{Array List}, and set its \textbf{Array Indexes} and related \textbf{ Dep Distance} as $[1,1]$ and $[-2, 0]$, respectively, we create two arrays $C[i+1][j+1]$ and $C[i+1-2][j+1]$. This constrains loop bounds to $1\le i\le M-1$ and $0\le j\le N-1$. Furthermore, if we introduce \textbf{Iterator  Bound} for $j<i$, loop bounds will change to $1\le i\le M-1$ and $0\le j<i$, ensuring the legality of array access.

During the configuration of loop properties, we employ a priority-based assignment to prevent contradictions. For example, array name property is constructed from \textbf{Read Dep}, \textbf{Write Dep} and \textbf{Array List}. In dependence-related parameters, \textbf{Write Dep} refers to WAW dependence, while \textbf{Read Dep} refers to WAR or RAW. \textbf{Read Dep} and \textbf{Write Dep} determine whether a statement contains dependence that contributes to the write to an array, while \textbf{Array List} provides an alternative for the write. We establish a priority rule that assigns higher priority to dependence-related parameters over \textbf{Array List} in array definition, preventing the generation of contradiction. 

However, the possibility of contradictions between loop parameters (e.g., between two dependence-related parameters) still remains. To address this issue, we propose a contradiction-check mechanism. For example, the information of the target array in WAW dependence,  controlled by \textbf{Write Dep}, is derived from the source array, while the information of the read to an array in WAR or RAW dependence, controlled by \textbf{Read Dep}, is derived from the write.
Therefore, creating circular dependence \cite{1130282272927219456} is not allowed, as tracing the source of dependence circularly will yield no usable information for array. After detecting any circular dependence, we drop the one associated with the statement that occurs later in the schedule to ensure legality.

% Although data dependence occurs when the same indexes of an array are accessed across different iterations, random generation of array access properties tends to produce code with little dependence, which are not suitable for diversity in loop transformations. Hence, we \textbf{Write Dep} specifies WAW dependence. \textbf{Read Dep} constrains dependence for WAR or RAW and the polarity of \textbf{Dep Distance} decides the final \textbf{Dep Type}, while positive number means WAR and negative number means RAW

Independent loop properties can be directly configured using parameters without the risk of contradictions.
The values of all parameters are randomized within a specified range to enable the possibility of triggering various loop transformations.

% 上段可删 enable the possibility of

% with the same name (e.g., loop depth, number of statements and dependence distance), or are grouped together to develop (e.g., $C[M][N]$ in \textbf{Array List} describe both array name $C$ and array size $M\times N$).

% 考虑是否还要大修
Finally, we convert the loop properties into SCoP of the code and complete the synthesis by integrating the SCoP with surrounding structure to form a complete program (Appendix \ref{10.2}).
After synthesizing the example codes, we use an optimization compiler and analyzers to obtain the optimized versions and extract data flow information for each example. The example codes, optimized versions and data flow information are stored as dataset for retrieval purpose.

\subsection{Retrieval} \label{4.2}

% Rag使用的匹配算法（bm25+额外信息+权重）
In the retrieval part, the target and example codes in dataset are extracted conforming to SCoP and evaluated for similarity. Existing retrieval methods primarily focus on text features, often overlooking loop properties. However, any subtle difference in SCoP, such as exchanging indexes in array access, can result in entirely different semantics and significantly impact the application of loop transformations.

% Due to the difference between code and natural language, we propose to construct a sparse retriever to deal with the information of loop properties.

To address this problem, we evaluate the similarity among SCoPs based on loop properties. However, not all loop properties are helpful for calculating similarity. For example, renaming all occurrences of an array's name in the SCoP has no effect on the application of loop transformations. Additionally, some loop properties can be derived from others. For instance, dependence occurs when the same indexes of an array are accessed across different iterations, and this can be expressed through loop schedules and array indexes.

% feature 和 properties 的关联性，见笔记4
On the other hand, observations from previous studies \cite{zhang2023automatic, wang2023selfconsistency} suggest that diversity in demonstrations may mitigate the effects of misleading guidance for LLMs. Therefore, to strike a balance between similarity and diversity, we introduce a loop-aware algorithm based on loop features extracted from loop properties of target and example SCoPs to calculate reward and penalty scores. The loop features encompass loop schedules and array indexes, which are closely related to loop transformations. For instance, array indexes serve as indicators of loop interchange when they are not aligned with the order of iterators in loop nests, and deserve reward scores.

%记得简介一下bm25
%公式要反着写，从细节到总体
The penalty \textbf{S}core for statements \textbf{M}ismatch ($S_{M}$) is calculated when the example and target SCoPs have a different number of statements, as given by:
\begin{equation}
S_{M} = |{NS}^{T} - {NS}^{E}| \times \sum_{0\le i\le N_{F}}{WP}_{i}
\end{equation}
where ${NS}^{T}$ and ${NS}^{E}$ refer to the \textbf{N}umber of \textbf{S}tatements in \textbf{T}arget and \textbf{E}xample SCoP respectively, $N_{F}$ denotes the \textbf{N}umber of types for \textbf{F}eature  and ${WP}_{i}$ represents \textbf{W}eight for \textbf{P}enalty on the $i$-th feature.

The \textbf{S}core for \textbf{F}eatures ($S_{F}$) contains both reward and penalty when statement matches. The \textbf{R}eward score ($R$) is designed to capture the intersection of features between the target and example SCoPs, as given by:
\begin{equation}
R_{i,j} = Count(F^{T}_{i,j} \cap F^{E}_{i,j}) \times WR_{j}
\end{equation}
where $R_{i,j}$ is the \textbf{R}eward score for the $j$-th feature in the $i$-th statement, $F^{T}_{i,j}$ and $F^{E}_{i,j}$ represent the information of $j$-th \textbf{F}eature in the $i$-th statement for \textbf{T}arget and \textbf{E}xample SCoP, and ${WR}_{j}$ indicates \textbf{W}eights for \textbf{R}eward of the $j$-th feature.

The \textbf{P}enalty score ($P$) is designed to address the redundancy within the SCoP of example code, aiming to deter the retrieval of example SCoP that employs inappropriate loop transformations. 
The score is calculated as follows:
\begin{equation}
P_{i,j} = (Count(F^{T}_{i,j} \cap F^{E}_{i,j}) - {NF}^{E}_{i,j}) \times WP_{j}
\end{equation}
where $P_{i,j}$ is the \textbf{P}enalty score for the $j$-th feature in the $i$-th statement and ${NF}^{E}_{i,j}$ denotes the \textbf{N}umber of the $j$-th \textbf{F}eature in the $i$-th statement for \textbf{E}xample SCoP. $Count(F^{T} \cap F^{E}) - {NF}^{E}$ calculates the number of unmatched features in example SCoP compared to target SCoP. This implies that a penalty should be applied when the example SCoP contains more features than target SCoP, as this may lead to unwanted demonstrations of loop transformations. Conversely, while the opposite situation (fewer features in the example SCoP) results in less diverse demonstrations, it is considered less harmful than inappropriate demonstrations. Therefore, we apply penalties only for cases where the example SCoP has more features.

The score $R$ and $P$ are calculated and grouped into \textbf{S}core for \textbf{F}eatures $S_{F}$ as follows: 
\begin{equation}
S_{F} = \sum_{0\le i\le min({NS}^{T}, {NS}^{E}), 0\le j\le N_{F}}{\frac{ R_{i,j} - P_{i,j} }{{NF}^{T}_{i,j}}}
\end{equation}
where ${NF}^{T}_{i,j}$ denotes the \textbf{N}umber of the $j$-th \textbf{F}eature in the $i$-th statement for \textbf{T}arget SCoP. $S_{F}$ is normalized by ${NF}^{T}$ to use the proportion of relative numbers rather than absolute numbers for calculating the similarity of loop features, and to examine the similarity of example compared to target.

After calculating the scores for loop properties, the overall Loop-Awared Score (LAScore) is defined as follows:
\begin{equation}
LAScore = S_{B} + S_{W} = S_{B} + \frac{ S_{F} - S_{M} }{{NS}^{T}}
\end{equation}
where $S_{B}$ denotes the base score about text similarity (e.g., BM25 \cite{INR-019}) which ensures syntactic robustness, and $S_{W}$ denotes the weighted score for high-level semantic features. $S_{M}$ and $S_{F}$ are normalized by ${NS}^{T}$ to eliminate the influence of the number of statements.

% \begin{equation}
% S_{F} = 
% \begin{cases}
% \sum_{N_{E}}{\sum_{N_{F}}{W_{F} \times \frac{(F_{T} \cap F_{E})\times (W_{R} + W_{P}) - N_{F{E}} \times W_{P}}{N_{F_{T}}}}} & \text{if} F_{T} \neq None \\
% 0   & \text{if} F_{T} = None
% \end{cases}
% \end{equation}

Sparse retrieval algorithms demonstrate great robustness when applied to code retrieval compared to traditional text retrieval, making them particularly advantageous for domain adaptations \cite{thakur2021beir} and enhancing retrieval efficiency. We choose BM25 as a representative of them to calculate the base similarity score for LAScore.
After computing the LAScore, we select example SCoPs with the top-N scores. These retrieved example SCoPs, along with their optimized versions, are grouped to prompt LLMs.

% \begin{algorithm}[!tb]
% \caption{Loop-aware Retrieval Algorithm}
% \label{algo:2}
% \small
% \SetKw{bflen}{Length}
% \KwIn{Example, Code, BaseScore, Weights, Penalties}
% \KwOut{Score}
% \textit{Score, QuerySize, DocSize} = \textit{BaseScore},  Length(\textit{Code}), Length(\textit{Example})\;
% \textit{SumPenalties} = Sum(\textit{Weights} $\times$ \textit{Penalties})\; \label{algo:2.2}

% \If{\textit{QuerySize - DocSize} $>$ 0}{
%     \textit{Score} -= (\textit{QuerySize - DocSize}) $\times$ \textit{SumPenalties}\; \label{algo:2.4}
% }

% \For{i = $0$ \KwTo \textit{DocSize} - $1$}{
%     \If{$i \geq$ \textit{QuerySize}}{ \label{algo:2.6}
%         \textit{Score} -= \textit{SumPenalties}\;
%         \textbf{continue}\; \label{algo:2.8}
%     }
    
%     \For{$j$ = $0$ \KwTo \bflen$($\textit{Code}$[i]$.\textit{infos}$)$ - $1$}{
%         \If{\textit{Code}$[i]$.\textit{infos}$[j]$ $\neq$ \textit{None}}{
%             \textit{Score} += \textbf{getLAScore}$($\textit{Weights}$[j]$, Penalties$[j]$, \textit{Code}$[i]$.\textit{infos}$[j]$, \textit{Example}$[i]$.\textit{infos}$[j]$$)$\; \label{algo:2.11}
%         }
%     }
% }

% \textit{Score} /= \textit{QuerySize}\; \label{algo:2.12}

% \Return \textit{Score} \label{algo:2.13}
% \end{algorithm}

\subsection{Feedback-based Iterative Generation} \label{4.3}

% Llm的prompt设计（规则引导和问题避免）
% Llm反馈机制（多轮多结果+编译、执行、时间排序+综合单一生成）
% The mechanism of feedback and iteration can significantly enhance the generation of LLMs by enabling LLMs to take more reasoning steps which further towards solving the failure and coming to a profitable answer \cite{grubisic2024compilergeneratedfeedbacklarge,NEURIPS2022_9d560961}

LLMs do not always generate the best output on their first try and often require feedback mechanism to learn from their mistakes and improve over time \cite{NEURIPS2023_91edff07}. Therefore, in code generation part, LOOPRAG is designed to provide informative feedback, iteratively guiding LLMs to generate correct and efficient SCoPs.

% 现有分类并address performance
Following the previous work \cite{agarwal2024structuredcoderepresentationsenable, gao2024search}, we categorize the common issues in code generation into five types based on correctness and performance:  compilation error (CE), incorrect answer (IA), runtime error (RE), execution timeout (ET) and inefficient code (IC).

% 所以，我们提出分别通过三个反馈机制

We present three types of feedback to guide LLMs: compilation results, testing results, and performance rankings. Compilation results help LLMs fix syntax errors (CE). For execution errors (i.e., IA, RE, and ET), testing results are provided to enable LLMs to refine the code. For inefficient code (IC), inspired by learning to rank \cite{INR-016}, LOOPRAG ranks all passing codes based on their performance, providing hints for LLMs to generate more efficient code.

These feedback mechanisms are integrated into a feedback-based iterative generation procedure, which consists of four steps:

\begin{enumerate}[Step 1:]
\setlength{\itemsep}{0pt}
    \item \label{enum:1.1} Construct demonstration with example SCoPs retrieved for target SCoP and corresponding optimized versions. Generate optimized SCoPs using prompts with demonstrations, integrate them into to programs and compile the resulting codes;
    \item \label{enum:1.2} Regenerate the codes that fail to pass the test in Step \ref{enum:1.1} using the compilation results. Test the regenerated codes along with the codes that passed the test in Step \ref{enum:1.1} for correctness. Execute and evaluate the passing codes to rank their performance;
    \item \label{enum:1.3} Incorporate testing results and performance rankings to guide LLMs in learning from execution errors and effective optimizations, and initiate a subsequent round of code generation;
    \item \label{enum:1.4} Repeat the compilation, generation and testing procedure. Compare the performance results and select the most optimal one as the final output of the generation procedure.
\end{enumerate}

% 4step说明->feedback的目的和原因（说明feedback的重要性，从错误中学习；四种错误，分别通过compile和performance实现并反馈）->而具体的测试怎么做的？（代码等价性验证）

It is challenging to verify whether loop transformations have preserved the original functionality, as the problem of proving semantic equivalence between two programs is a long-standing and undecidable problem \cite{dramko2025fastfinegrainedequivalencechecking}.
Several approaches have been proposed, such as code invariant \cite{pmlr-v202-pei23a}, formal verification \cite{10.1145/3453483.3454030} and checkpoints comparison \cite{10.1145/3624062.3624140}.
However, these methods struggle to scale to real-world programs \cite{gupta2018effective} (e.g., functions involving two-dimensional arrays \cite{10.1145/3314221.3314596}). Pragmatically, we adopt approaches from compilers testing for equivalence checking \cite{10.1145/3363562}, and the testing results are combined with performance rankings to guide LLMs.

% Checksum
% Element-wise
% Cover——以原始输入为seed、基于LLM进行随机初始化、采用分支覆盖率100%或饱和作为测试充分性引导
Specifically, we employ LLM-assisted mutation method to generate and diversify test inputs.
First, GPT-4 is utilized to learn the input structures (e.g., data types and dimensions of arrays) and functionalities of the ground-truth program, and then to generate a set of initialization functions as seed inputs \cite{NEURIPS2023_43e9d647}.
Using these seed inputs, LOOPRAG performs value-based, operator-based and statement-based mutations to efficiently create diverse test inputs.
% These seed functions are then mutated by directly changing the number of result arrays (e.g., reorder, exchange and apply affine transformation to the elements in arrays) or indirectly changing calculation methods (e.g., add conditional branch) to create diverse test cases.

A coverage-guided testing approach is designed to achieve high branch coverage. 
When the branch coverage reaches 100\% or saturates, LOOPRAG considers the program to be adequately tested. 
By employing coverage-guided testing, the number of test cases has been significantly reduced, from an average of 500+ to just 25 for each program in the benchmark suites.
% 要不要讲500哪来的？
% while others reach for 2.03 \cite{zhou2023docprompting}).

We introduce the differential testing approach as the oracle to cross-check the output of the ground-truth and LLM-generated codes.
Two types of testing are used: checksum testing and element-wise testing. 
The former acts as a quick filter to match checksum results from the two codes, while the latter performs an element-wise comparison for equivalence checking.
% in element-wise testing by applying different array initialization.

\section{Implementation} \label{5}

\begin{figure}[tb]
    \centering
    \includegraphics[width=\linewidth]{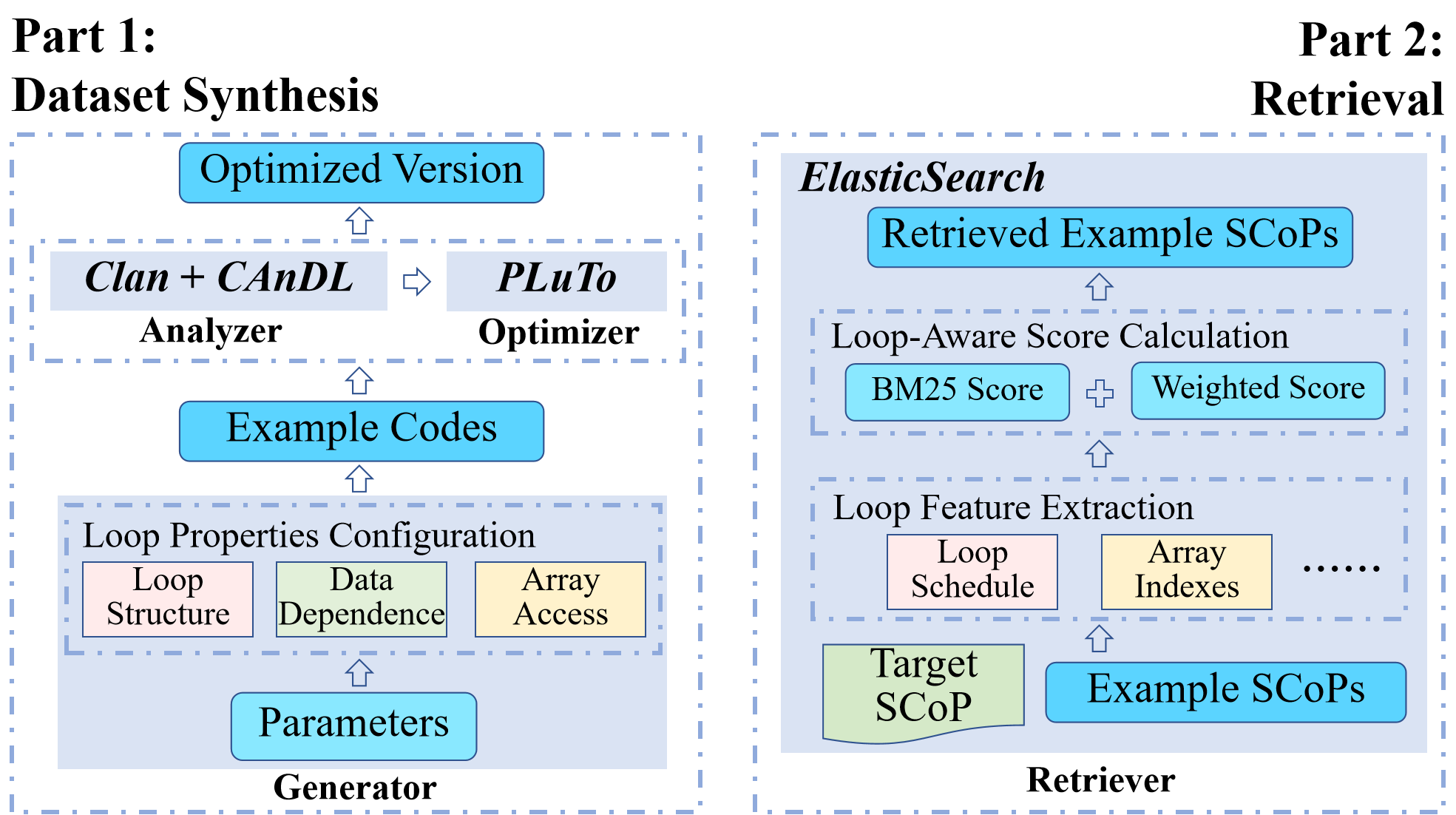}
    \caption{Implementation of Code Synthesis and Retrieval.}
    \label{fig:5}
\end{figure}

% 先说param-driven random生成examples
As shown in Figure \ref{fig:5}, LOOPRAG employs a parameter-driven method in code generator to randomly set parameters, configure loop properties, and synthesize example codes while ensuring their legality. We employ Clan 0.8.0 and CAnDL 0.6.2 as analyzer. The options for Clan and CAnDL are with default. 
We select PLuTo as the optimization compiler for example codes due to its reliable optimization and ease of manipulation. We modify PLuTo 0.11.4 to fulfill our need for information extraction about loop properties, and add function for global parameters specification, which reduces redundant conditional branches in optimized version. The option of PLuTo used in LOOPRAG is -q -custom-context -tile -parallel -nocloogbacktrack.
Clan and CAnDL are used to extract data flow information, and PLuTo is used to generate optimized versions of example codes. SCoP in code is surrounded by directives $\#pragma\ scop$ and $\#pragma\ endscop$ to indicate the region that LOOPRAG and PLuTo will detect and optimize.

In our retriever, Elasticsearch \cite{10.5555/2904394} 7.13.2 is deployed as the data storage and retrieval platform. LOOPRAG proposes a loop-aware retrieval algorithm based on BM25 algorithm, focusing on extracting loop features in loop properties from target and example SCoPs (Appendix \ref{10.3}). We set the number of retrieved example SCoPs N to ten and randomly select three of them, along with their optimized versions, as demonstrations to prompt LLMs. We provide specific prompt rules to guide LLMs in learning from feedback and generating SCoPs iteratively (Appendix \ref{10.4}). The SCoPs are then completed to form the codes. The value of the number of generated codes K is set to 7.

% In testing procedure, for the potential accidents in initialization, we check all elements in arrays and replace all errors (e.g., NaN) and outliers (e.g., inf) by normalize the value of elements to suitable bound (e.g., -1 to 1) to reduce numerical overflow happened in further testing. 

For three types of feedback, we obtain the compilation results from compilers (e.g., GCC). The testing results are derived from a combination of seed inputs mutation, branch coverage and differential testing, estimating the output arrays at the end of the SCoP and measuring the execution time of the whole SCoP. The seed inputs are generated using GPT-4, and we use gcov tool to test branch coverage. We utilize $omp\_get\_wtime$ from OpenMP instead of $rtclock$ or $rdtsc$ instruction for precision consideration.

% since $rtclock$ based on the underlying hardware and operating system and $rdtsc$ count CPU cycles, which may affect by other system operation and are not applicable for parallelism usage.

% The code that not passes will be checked again for the possible incorrect error of floating point numbers during the testing for output equivalence. 

\section{Evaluation} \label{6}
We design our experiments to address the following research questions:
\begin{enumerate}
    \item How effective is LOOPRAG in enhancing loop optimization?
    \item Can LOOPRAG surpass its optimization demonstration source PLuTo?
    \item What are the contributions of different modules to LOOPRAG?
    \item How does LOOPRAG help improve the quality of loop optimization?
\end{enumerate}
% 共5.5 —— 10.25

\subsection{Experiment Setup} \label{6.1}

%要解释下LORE从各个数据集收集，多样性更大
%要说主要GCC

\textbf{Hardware and Software Specifications.}
Our work focuses on CPU targets. All experiments are conducted on a Linux server (64-bit Ubuntu 20.04.4) with two 24-core AMD EPYC 7352 CPUs @ 2.3GHz, 96 thread; eight NVIDIA RTX 4090 GPUs; and 256GB RAM, DDR4 @ 3.2GHz. We utilize our parameter-driven method to synthesize 135,364 example codes \footnote{https://anonymous.4open.science/r/Dataset-for-LOOPRAG-48C5}. For base LLMs, we use DeepSeek-V3 (deepseek-v3-0324) and GPT-4 (gpt-4o-2024-08-06), and configure temperature of LLMs to 0 to ensure deterministic behavior during code generation. We use GCC 15.1.0 as the base compiler with option -O3, -fopenmp. 

\textbf{Benchmarks.}
We evaluate LOOPRAG on three benchmark suites: PolyBench \cite{pouchet2023PolyBench}, TSVC \cite{44642} and LORE \cite{8167779}. PolyBench includes 30 numerical computations extracted from real-world applications and is widely used to evaluate the performance of loop optimization techniques in compilers. TSVC, consisting of 149 \emph{for} loops, is designed to assess compiler optimizations related to vectorization and parallelism. LORE contains a large amount of \emph{for} loop nests extracted from benchmark suites, libraries and real-world applications, and is used for evaluating compilers.

We extract benchmarks that satisfy the requirements of SCoP from these test suites. We manually add directives for TSVC and LORE, which allow analyzers (e.g., Clan) to identify SCoP in each benchmark. These directives have existed in PolyBench. In total, PolyBench consists of 30 benchmarks, while TSVC and LORE consist of 84 and 49 benchmarks, respectively. As for the value of parameters, we use option EXTRALARGE\_DATASET for PolyBench, apply default settings for TSVC, and manually set value for parameters in LORE. We add \emph{\_\_attribute\_\_((pure))} qualifier to the \emph{dummy} function in TSVC to ensure correct SCoP detection by Clang-Polly, whereas GCC-Graphite fails to recognize the SCoP in TSVC due to the opacity of the \emph{dummy} function. Details regarding the SCoP handling strategy in our evaluation are provided in Appendix \ref{10.8}

We compare the synthesized codes with codes in benchmark suites, confirming that no identical code exists. The output arrays verified at the end of program during testing are predefined in PolyBench and TSVC. As for LORE, we manually identify some functionally relevant arrays (e.g., the only write to array in SCoP).

\textbf{Baseline.}
We evaluate LOOPRAG against four categories of baselines: four leading compilers, two representative LLMs, two LLM-based models specifically designed for code optimization, and PLuTo (the learning objective of LOOPRAG).

For the baseline compilers, we employ Polly \cite{grosser2012polly} with Clang 20.1.4, Graphite \cite{pop2006graphite} with GCC 15.1.0, ICX 2024.2.0, and Perspective \footnote{https://github.com/PrincetonUniversity/cpf} \cite{apostolakis2020perspective} with Clang 9.0.2. Polly and Graphite integrate the polyhedral model into production compilers to enable loop parallelization, whereas Perspective represents a state-of-the-art (SOTA) system that applies speculative techniques for automatic loop parallelization.
The compilation options are set as -O3 -fopenmp -mllvm -polly -mllvm -polly-parallel -mllvm -polly-tiling for Clang-Polly,  -O3 -fopenmp -floop-nest-optimize -floop-parallelize-all for GCC-Graphite, and -O3 -qopenmp -xHost for ICX. $TRANSFORM\_TIMEOUT$ and $PROFILE\_TIMEOUT$ in Perspective are set to 600s. PLuTo 0.11.4 uses option -tile -parallel -nocloogbacktrack.

For LLMs, we compare LOOPRAG with: DeepSeek-V3 (deepseek-v3-0324) and GPT-4 (gpt-4o-2024-08-06). These two LLMs are used as both base LLMs of LOOPRAG and the baselines in comparison. They represent code-specialized and general-purpose language models, respectively. We access both models through their official APIs using instruction prompting.

Regarding LLM-based models, we select PCAOT \cite{rosas2024aioptimizecodecomparative} and LLM-Vectorizer \cite{taneja2024llmvectorizerllmbasedverifiedloop} as they represent the SOTA methods for code optimization. PCAOT prompts LLMs to generate optimized code for parallel computing using instruction prompting and chain-of-thought (COT) strategies. LLM-Vectorizer explores the capabilities of LLMs and AI-based agents to perform efficient loop vectorization optimizations. Since neither approach has released their software, we report the results directly from their respective papers.

\textbf{Metrics.}
We evaluate the generated codes using three metrics, including pass@k \cite{NEURIPS2019_7298332f}, performance speedup, and the percentage of faster codes. The pass@k metric measures the proportion that at least one of the top-K generated codes passes a set of correctness tests.

The speedup metric is calculated as arithmetic mean of the performance improvement (measured over five runs after the first attempt) by comparing the execution time of SCoP in the optimized code to that of the original code. All codes are compiled using GCC in idle test environment by default.

The percentage of faster codes reflects the proportion of codes with higher speedup, thereby avoiding the instability associated with a single average speedup metric.

We enforce a performance limit of 120 seconds for SCoP in codes as a threshold of successfully applying efficient optimizations for LOOPRAG. This threshold is expanded to 600s for baselines. The speedup of failure code is set to 0. The average speedup metric is calculated by excluding outliers (e.g., speedup greater than 600$\times$) to reduce standard deviation error.

\subsection{How effective is LOOPRAG in enhancing loop optimization?}

\subsubsection{Performance Comparison with Compilers} \label{6.2.1}
\begin{table}[thb]\centering
    \caption{Pass@k and speedups achieved by LOOPRAG compared to baseline compilers. \emph{LD-GCC} and \emph{LG-GCC} denote LOOPRAG configured with GCC as the base compiler, using DeepSeek and GPT-4 as the base LLMs, respectively. \emph{LD-Clang} and \emph{LG-Clang} denote LOOPRAG configured with Clang as the base compiler, using DeepSeek and GPT-4 as the base LLMs, respectively. \emph{LD-ICX} and \emph{LG-ICX} denote LOOPRAG configured with ICX as the base compiler, using DeepSeek and GPT-4 as the base LLMs, respectively.} %
    \resizebox{0.48\textwidth}{!}{
    \fontsize{16}{20}\selectfont
    \begin{tabular}{{c}*{2}{c}*{2}{c}*{2}{c}}
        \toprule[1.5pt]
        \multirow{2.5}{*}{Compilers} & \multicolumn{2}{c}{PolyBench} & \multicolumn{2}{c}{TSVC} & \multicolumn{2}{c}{LORE} \\
        \cmidrule{2-7}
          & Pass@k  & Speedup  & Pass@k  & Speedup  & Pass@k  & Speedup  \\
        \hline
        \emph{LD-GCC} & 70.00 & \textbf{23.97} & \textbf{94.05} & \textbf{32.66} & 86.71 & \textbf{20.44} \\ 
        \emph{LG-GCC} & 70.00 & 14.58 & 84.52 & 31.35 & 87.76 & 14.86 \\
        Graphite & \textbf{96.67} & 0.99 & - & - & \textbf{100} & 1.03 \\
        \midrule
        \emph{LD-Clang} & 66.67 & 8.62 & 94.05 & 15.82 & 85.71 & \textbf{6.49} \\
         \emph{LG-Clang} & 70.00 & 14.06 & 84.52 & 24.00 & 87.76 & 5.01 \\
         Polly & \textbf{93.33} & \textbf{14.62} & \textbf{100.00} & \textbf{27.54} & \textbf{100.00} & 1.05 \\
         Perspective & 26.67 & 1.79 & - & - & 61.22 & 3.07\\
        \midrule
        \emph{LD-ICX} & \textbf{66.67} & 9.63 & \textbf{94.05} & 12.30 & 85.71 & \textbf{5.63} \\
        \emph{LG-ICX} & \textbf{66.67} & \textbf{12.38} & 83.33 & \textbf{23.28} & \textbf{87.76} & 4.90 \\
        \bottomrule[1.5pt]
    \end{tabular}
    }
    \label{tab:2}
\end{table}

To evaluate the performance of LOOPRAG in improving code efficiency, we first compare LOOPRAG with four compilers, including Clang-Polly, GCC-Graphite, ICX and Perspective. The speedups are calculated by comparing performance of the optimized code to original code on each compiler.

Since each TSVC kernel contains a \emph{dummy} function call, SCoP detection in Graphite fails unless the function is explicitly guaranteed to be side-effect free. Therefore, we exclude Graphite from the TSVC comparison to ensure fairness. In addition, Perspective cannot handle TSVC codes with large iteration sizes, as the outermost loops require 100,000 iterations, causing timeouts during the profiling process. Consequently, we also omit TSVC results for Perspective.

% (全部采用looprag相比graphite，polly等的加速比，而不是相比GCC)
As shown in Table \ref{tab:2}, LOOPRAG achieves significant performance gains over Graphite and ICX across the benchmark suites. Specifically, compared to Graphite, LOOPRAG achieves average speedups of 19.47$\times$ (i.e., $(23.97/0.99 + 14.58/0.99)/2$) on PolyBench and 17.14$\times$ on LORE. Against ICX, LOOPRAG delivers average speedups of 18.18$\times$, 27.97$\times$, and 12.67$\times$ on PolyBench, TSVC and LORE, respectively. 

% abstract 两个平均（compilers和base llm） introduction 三个平均（多pcaot和llm-vectorizer的平均）
% 说完总平均，看回table，gpt表现出了可比相近结果，所以还是不错
As for Polly, LOOPRAG exhibits varying advantages across the benchmark suites, with average speedups of 0.78$\times$, 0.72$\times$, and 5.48$\times$ on PolyBench, TSVC, and LORE, respectively. In particular, we observe that \emph{LG-Clang} outperforms \emph{LD-Clang} and achieves performance comparable to Polly on PolyBench and TSVC. This suggests that \emph{LG-Clang}, which employs GPT-4 as the base LLM, may cooperate more effectively with the the base compiler Clang in performing code optimization.

Compared to Perspective, LOOPRAG outperforms it in both pass@k and speedup. The average speedup improvements (both on Clang-9) reach $6.34\times$ and $1.87\times$ on PolyBench and LORE, respectively. Upon inspecting the failure cases of Perspective, we find that it relies on a delicate and time-consuming profiling and optimization process. For example, the TSVC code leads to excessive timeouts, and the code with varying complexity in PolyBench and LORE makes it prone to failures during analysis.

% perspective 分析过后，他需要长分析时间，所以不佳

% Among the results above, LOOPRAG exceeds Graphite on all benchmark suites and Polly on LORE, since these compilers hardly success to find opportunity to optimize the codes. Perspective shows limitation in time consumption for applying optimization. In general, LOOPRAG presents its advantage for adaptability and convenience in code optimization on SCoP, but still needs improvements on both correctness and performance improvement in the future.

\begin{figure}[thb]
    \centering
    \includegraphics[width=0.48\textwidth]{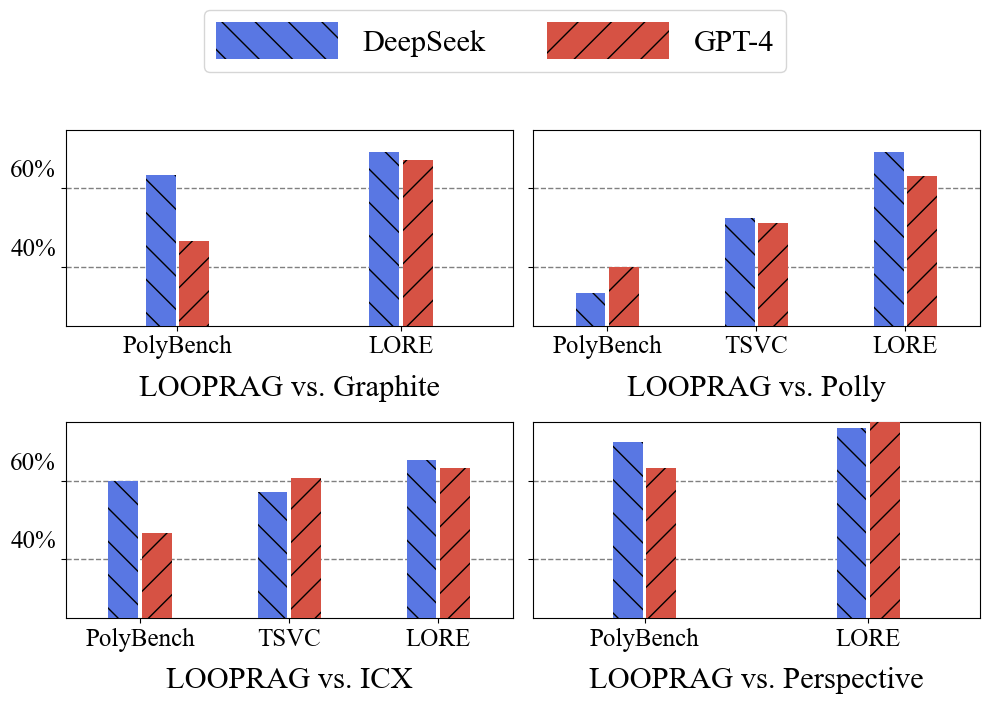}
    \caption{Percentage of faster codes generated by LOOPRAG compared to four compilers.}
    \label{fig:6}
\end{figure}

Figure \ref{fig:6} illustrates the percentage of faster codes generated by LOOPRAG compared to four compilers. We observe that LOOPRAG produces more than 40\% of faster codes than Graphite, ICX and Perspective on all available benchmark suites, even touching 80\% of faster codes on LORE. As for Polly, LOOPRAG has more than 50\% of faster codes on TSVC and LORE, but achieves much less on PolyBench, demonstrating the power of Polly on codes in complexity.

Besides, LOOPRAG produces more faster codes for TSVC and LORE than for PolyBench. A potential reason is that the codes in PolyBench exhibit more complex loop properties compared to those in TSVC and LORE. As a result, it is more challenging for LOOPRAG to perform effective loop optimizations on PolyBench.

% Although there exists difference in value of average speedup for different compilers, DeepSeek has little difference for the percentage of faster codes on all three benchmarks. GPT-4 shows completely identical result on PolyBench, but different on TSVC and LORE, which means GPT-4 is more sensitive for compilers on benchmarks with less complexity.
In general, compared with the polyhedral compilers (Polly and Graphite), the general-purpose compiler ICX, and the speculation-based compiler Perspective, LOOPRAG demonstrates superior performance and adaptability across all benchmark suites.

\subsubsection{Comparison with LLM-based Methods} \label{6.2.2}

\begin{table}[thb]\centering
    \caption{Pass@k and speedups achieved by LOOPRAG compared to LLM-based methods. The value of $k$ is set to 7 for LOOPRAG, base LLMs and LLM-Vectorizer, while PCAOT uses 12.} %
    \resizebox{0.48\textwidth}{!}{
    \fontsize{16}{20}\selectfont
    \begin{tabular}{*{2}{c}*{2}{c}*{2}{c}*{2}{c}}
        \toprule[1.5pt]
        \multirow{2.5}{*}{Methods} & \multirow{2.5}{*}{LLMs} & \multicolumn{2}{c}{PolyBench} & \multicolumn{2}{c}{TSVC} & \multicolumn{2}{c}{LORE} \\
        \cmidrule{3-8}
         &  & Pass@k  & Speedup  & Pass@k  & Speedup  & Pass@k  & Speedup  \\
        \hline
        \multirow{2}{*}{LOOPRAG} & DeepSeek & \textbf{70.00} & \textbf{23.97} & 94.05 & \textbf{32.66} & 86.71 & \textbf{20.44} \\
        & GPT-4 & \textbf{70.00} & 14.58 & 84.52 & 31.35 & 87.76 & 14.86 \\
        \midrule
        \multirow{2}{*}{Base LLMs} & DeepSeek & 66.67 & 1.61 & \textbf{97.62} & 6.75 & \textbf{91.84} & 1.72 \\ 
        & GPT-4 & \textbf{70.00} & 1.61 & 92.86 & 4.91 & 85.71 & 1.60 \\
        \midrule
		\multirow{2}{*}{PCAOT} & GPT-4 & 65.35 & 1.80 & - & - & - & - \\
        & CLLama-70B & 63.35 & 2.26 & - & - & - & - \\
        \midrule
		LLM-Vectorizer & GPT-4 & - & - & 68.00 & 5.25 & - & - \\
        \bottomrule[1.5pt]
    \end{tabular}
    }
    \label{tab:1}
\end{table}

% Although LOOPRAG outperforms industry compilers on performance, it reveals a possible weakness on correctness, and its optimization accomplishment still remains a challenge that it might comes from the ability of LLMs themselves but not LOOPRAG method. 

As for LLM-based methods, we optimize the original code using LOOPRAG and each LLM-based baseline, respectively, and then compare the results after compilation and testing.

As shown in Table \ref{tab:1}, compared to two representative base LLMs (i.e., DeepSeek and GPT-4), LOOPRAG demonstrates high pass@k performance while achieving significantly higher efficiency in terms of speedup. Specifically, LOOPRAG achieves average speedup improvements of 11.97$\times$ (i.e., $(23.97/1.61 + 14.58/1.61)/2$), 5.61$\times$ and 11.59$\times$ for PolyBench, TSVC and LORE, respectively. These results highlight the superiority of LOOPRAG in performing effective loop optimizations.

% The pass@k metric of these optimized codes reaches 60\%, 85.71\% and 91.83\% when using DeepSeek, and 60.00\%, 89.29\% and 85.71\% when using GPT-4. These results are consistent across all three compilers, since the choice of compiler does not influence code correctness. We observe that GPT-4 is more effective in preserving semantics while generating optimized codes.

Comparing the results on DeepSeek and GPT-4, we observe that LOOPRAG based on DeepSeek (deepseek-v3-0324) achieves better speedups than GPT-4 (gpt-4o-2024-08-06), largely due to differences in the release time of the underlying LLMs. Specifically, DeepSeek-v2.5 delivers lower speedups than GPT-4 on Polybench. It indicates that LOOPRAG can help LLMs build a deeper understanding of complex code logic and context, and remains effective for code optimization on SCoP with newer base LLMs.

Regarding the SOTA LLM-based models, we compare LOOPRAG with PCAOT and LLM-Vectorizer. Two key observations can be drawn from Table \ref{tab:1}. First, compared to PCAOT, LOOPRAG achieves comparable pass@k performance while delivering an average speedup improvement of 8.10$\times$ for PolyBench. Second, in comparison to LLM-Vectorizer, LOOPRAG achieves approximately 16.52\% higher pass@k performance, along with average speedup improvements of 5.97$\times$ for TSVC. These results highlight the effectiveness of LOOPRAG in both efficiency and correctness.

\begin{figure}[thb]
    \centering
    \includegraphics[width=0.36\textwidth]{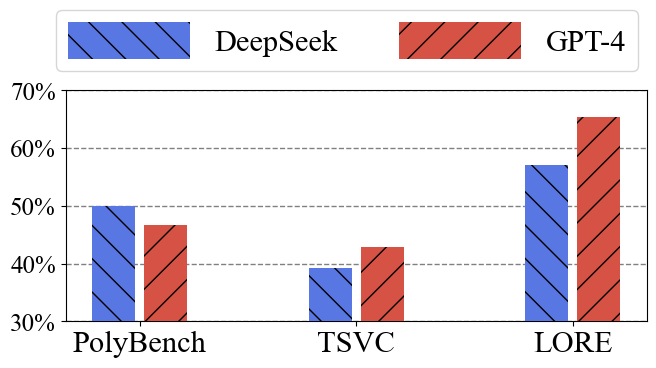}
    \caption{Percentage of faster codes generated by LOOPRAG compared to base LLMs.}
    \label{fig:6.1.1}
\end{figure}

In Figure \ref{fig:6.1.1}, although LOOPRAG generates more optimized codes for LORE (about 60\%) than for PolyBench (about 50\%), their performance improvements are very close, at 11.97$\times$ and 11.59$\times$, respectively. This observation suggests that the codes in PolyBench offer more opportunities for LOOPRAG to perform efficient optimizations compared to LORE. More details are discussed in Appendix \ref{10.5}.

% The speedup metric for benchmarks and LLMs presents different situations. Due to the difference in complexity of benchmarks, both the average speedup and percentage of faster codes metrics for PolyBench is lower than others. GPT-4 outperforms DeepSeek on PolyBench for both average speedup and the percentage of faster codes metrics and the percentage of faster codes metric on LORE, which further explains the power of GPT-4.

%By investigating the difference between benchmarks, we observe that the speedup for codes in TSVC is relatively lower in general. However, its speedup in \ref{tab:1} is higher than others while its percentage of faster codes in Figure \ref{fig:5} is not significantly prominent. It means that several optimized codes in TSVC reach extremely high speedups, distorting the metric of average speedup. For example, code $s114$ and $s1232$ reach 214.54$\times$ and 248.30$\times$ for LOOPRAG using DeepSeek, while GPT-4 gets 6.32$\times$ and 7.09$\times$. These codes optimized through DeepSeek fully utilized loop tiling and loop interchange to achieve parallelism and better data locality, lead to high speedup in TSVC. Further analysis of loop transformations applied in optimized codes for LOOPRAG are presented in §\ref{6.4}.

\subsection{Can LOOPRAG surpass its optimization demonstration source PLuTo?} \label{6.4}
\begin{table}[thb]\centering
    \caption{Pass@k and speedups achieved by LOOPRAG compared to PLuTo.} %
    \resizebox{0.48\textwidth}{!}{
    \fontsize{16}{20}\selectfont
    \begin{tabular}{*{2}{c}*{2}{c}*{2}{c}*{2}{c}}
        \toprule[1.5pt]
        \multirow{2.5}{*}{Methods} & \multirow{2.5}{*}{LLMs} & \multicolumn{2}{c}{PolyBench} & \multicolumn{2}{c}{TSVC} & \multicolumn{2}{c}{LORE} \\
        \cmidrule{3-8}
         & & Pass@k  & Speedup  & Pass@k  & Speedup  & Pass@k  & Speedup  \\
        \hline
        \multirow{2}{*}{LOOPRAG} & DeepSeek & 70.00 & 23.97 & \textbf{94.05} & \textbf{32.66} & 85.71 & \textbf{20.44} \\ 
		& GPT-4 & 70.00 & 14.58 & 84.52 & 31.35 & 87.76 & 14.86 \\
        \midrule
        PLuTo & & \textbf{83.33} & \textbf{43.29} & \textbf{94.05} & 5.88 & \textbf{95.92} & 4.03 \\
        \bottomrule[1.5pt]
    \end{tabular}
    }
    \label{tab:3}
\end{table}

% PLuTo is unique compared to other baselines because it is utilized in generating optimized versions of example codes in dataset synthesis. 
LOOPRAG learns the strategy of applying the appropriate composition of loop transformations from the optimization compiler PLuTo. This raises the question: \textbf{Can LOOPRAG surpass its learning objective?}

Table \ref{tab:3} compares the optimized codes from LOOPRAG and PLuTo, while the only difference is the usage of optimizer. Our 600s execution timeout mechanism caused some PLuTo-optimized codes to fail, particularly with large arrays. LOOPRAG achieves average speedup improvements of \textbf{5.44$\times$ for TSVC and 4.38$\times$ for LORE. However, for PolyBench, LOOPRAG shows a reduced speedup of 0.45$\times$}, which aligns with the findings in PCAOT \cite{rosas2024aioptimizecodecomparative} that PLuTo's specialization in the polyhedral model \cite{feautrier2005automatic} makes it significantly more efficient for PolyBench. In contrast, for TSVC and LORE, PLuTo may apply unprofitable transformations (such as loop tiling on small iteration sizes) and incur overhead from excessive distribution and peeling, resulting in lower speedups.

In general, although LOOPRAG primarily learns and utilizes loop transformations from the optimized versions generated by PLuTo in its demonstrations, it does not limit the optimization capabilities of LLMs themselves. As a result, LOOPRAG can outperform PLuTo. For instance, LOOPRAG can generate more effective composition of loop transformations by considering additional optimization objectives, such as vectorization and data locality, and can employ auxiliary techniques like scalar renaming and reductions \cite{44642}, rather than relying solely on loop transformations.

\begin{figure}[thb]
    \centering
    \includegraphics[width=0.36\textwidth]{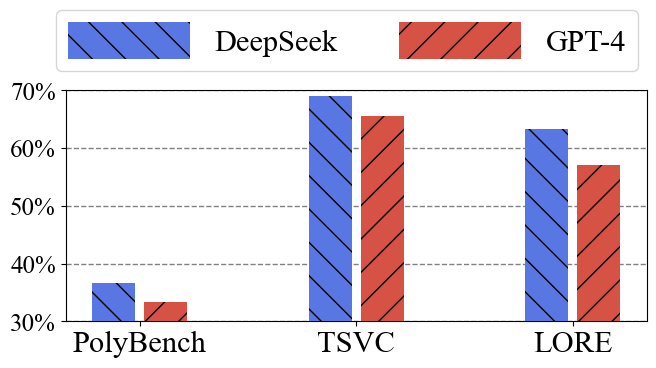}
    \caption{Percentage of faster codes generated by LOOPRAG compared to PLuTo.}
    \label{fig:7}
\end{figure}

Figure \ref{fig:7} illustrates the percentage of faster codes generated by LOOPRAG compared to PLuTo, further supporting the conclusions about PLuTo mentioned above. Although PLuTo outperforms LOOPRAG on more than 60\% of the codes in PolyBench, LOOPRAG achieves around 60\% more optimized codes in TSVC and LORE. These results demonstrate that LOOPRAG can serve as a practical code optimization tool and has greater potential for the development of diverse optimization techniques.

\subsection{What are the contributions of different modules to LOOPRAG?}\label{6.3}

To investigate the individual contributions of different components in LOOPRAG, we conduct three experiments to evaluate the impact of these methods.

\subsubsection{Example Codes Synthesis}\label{6.3.1}

% 实验目的->实验设计->实验结果（分析相对值）->原因分析
% ！！！注意不要重复实现原理及实现

To evaluate the impact of our parameter-driven method in dataset synthesis, we use COLA-Gen \cite{berezov_et_al:OASIcs.PARMA-DITAM.2022.3} as baseline to synthesize the same number of example codes (135,364) with LOOPRAG under its default settings (i.e., loop depth set to 2 and the number of array reads set to 1). We conduct three analytical studies to explore the distribution of loop properties in the example codes, the types of loop transformations triggered in the optimized versions, and the performance of the optimized codes.

\begin{figure}[thb] \centering
    \includegraphics[width=0.48\textwidth]{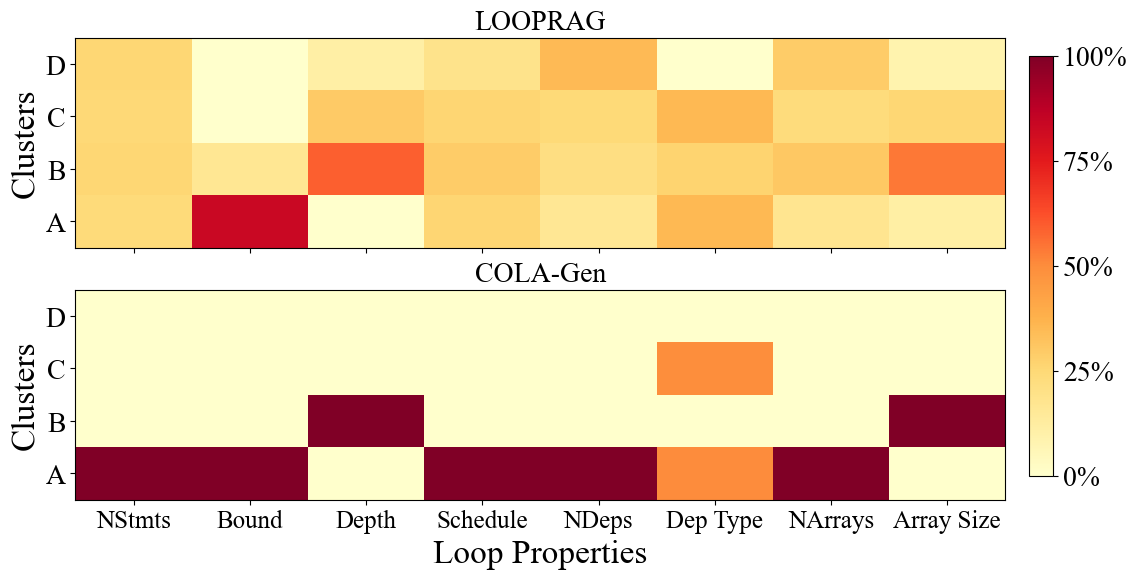}
    \caption{Distribution Percentage of loop properties in the SCoP of example codes synthesized by LOOPRAG and COLA-Gen. \textbf{NStmts}, \textbf{Bound}, \textbf{Depth}, \textbf{Schedule}, \textbf{NDeps}, \textbf{Dep Type}, \textbf{NArrays} and \textbf{Array Size} indicate loop properties for number of statements, loop bounds, loop depth, loop schedule, number of dependence, dependence type, number of arrays and array size, respectively.}
    \label{fig:8.1}
\end{figure}

Figure \ref{fig:8.1} displays the distribution of eight loop properties in the example codes. Each loop property is divided into four clusters based on its property value. For example, for \textbf{NDeps}, SCoPs containing 0, 1 or 2 dependence belong to cluster \textbf{A}, while SCoPs containing 3-5, 6-10 and at least 11 dependence belong to cluster \textbf{B}, \textbf{C} and \textbf{D}, respectively. We observe that in the example codes synthesized by COLA-Gen, all loop properties are highly concentrated in one or two clusters, whereas they are more uniformly distributed in those synthesized by LOOPRAG. This indicates that LOOPRAG, through its parameter-driven method, can synthesize more diverse example codes compared to COLA-Gen.

We use the optimization compiler PLuTo to produce the optimized versions and analyze the loop transformations triggered in these versions. Table \ref{tab:5} shows the differences in the types of loop transformations. The results demonstrate that the parameter-driven method in LOOPRAG synthesizes more diverse example codes, leading to a wider variety of loop transformations in the optimized versions.

\begin{table}[thb]\centering
    \caption{The differences in the types of loop transformations triggered in the optimized versions of example codes synthesized by LOOPRAG and COLA-Gen.}
    \resizebox{0.48\textwidth}{!}{
    \fontsize{16}{20}\selectfont
    \begin{tabular}{{c}*{6}{c}}
        \toprule[1.5pt]
        \multirow{2}{*}{Methods} & \multicolumn{6}{c}{Loop Transformations} \\
        \cmidrule(lr){2-7}
         & Tiling & Interchange & Skewing & Fusion & Distribution & Shifting \\
        \midrule
        LOOPRAG & $\checkmark$ & $\checkmark$ & $\checkmark$ & $\checkmark$ & $\checkmark$ & $\checkmark$ \\ 
        COLA-Gen & $\checkmark$ & $\checkmark$ & $\checkmark$ & $\times$ & $\times$ & $\times$ \\
        \bottomrule[1.5pt]
    \end{tabular}
    }
    \label{tab:5}
\end{table}

\begin{table}[thb]\centering
    \caption{Pass@k and speedups achieved by LOOPRAG compared to COLA-Gen.} %
    \resizebox{0.48\textwidth}{!}{
    \fontsize{16}{20}\selectfont
    \begin{tabular}{*{2}{c}*{2}{c}*{2}{c}*{2}{c}}
        \toprule[1.5pt]
        \multirow{2.5}{*}{Methods} & \multirow{2.5}{*}{LLMs} & \multicolumn{2}{c}{PolyBench} & \multicolumn{2}{c}{TSVC} & \multicolumn{2}{c}{LORE} \\
        \cmidrule{3-8}
         & & Pass@k  & Speedup  & Pass@k  & Speedup  & Pass@k  & Speedup  \\
        \hline
        \multirow{2}{*}{LOOPRAG} & DeepSeek & 70.00 & \textbf{23.97} & \textbf{94.05} & 32.66 & \textbf{85.71} & \textbf{20.44} \\ 
        & GPT-4 & 70.00 & 14.58 & 84.52 & 31.35 & 87.76 & 14.86 \\
        \midrule
        \multirow{2}{*}{COLA-Gen} & DeepSeek & \textbf{76.67} & 9.82 & \textbf{94.05} & \textbf{35.08} & \textbf{89.80} & 17.85 \\
		& GPT-4 & 60.00 & 2.82 & 82.14 & 12.94 & 83.67 & 11.44 \\
        \bottomrule[1.5pt]
    \end{tabular}
    }
    \label{tab:4}
\end{table}

\begin{figure}[thb]
    \centering
    \includegraphics[width=0.36\textwidth]{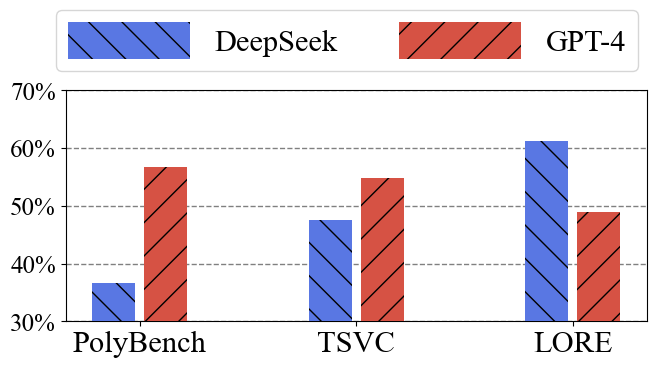}
    \caption{Percentage of faster codes generated by LOOPRAG compared to COLA-Gen.}
    \label{fig:8}
\end{figure}

Table \ref{tab:4} compares the performance of the optimized codes using example codes synthesized by LOOPRAG and COLA-Gen. Overall, LOOPRAG outperforms COLA-Gen in both pass@k and speedup. The average speedup improvements reach 3.81$\times$, 1.68$\times$ and 1.22$\times$ for PolyBench, TSVC and LORE, respectively. The results in Figure \ref{fig:8} also highlight the impact of parameter-driven method in dataset synthesis, which serves as a foundation for performing efficient code optimization.

% The improvement on average speedup reaches 3.62$\times$, 2.42$\times$ and 3.23$\times$ for PolyBench, TSVC and LORE respectively using DeepSeek, and 1.82$\times$, 0.46$\times$ and 1.17$\times$ respectively using GPT-4. 
% The lower improvement for GPT-4 shows that GPT-4 is less sensitive for provided examples in demonstrations, and the higher improvement for PolyBench emphasizes the importance of dataset towards loop transformation optimization on more complex benchmark.

% The percentage of faster codes for LOOPRAG compared to COLA-Gen in Figure \ref{fig:8} has similar difference as Figure \ref{tab:4} demonstrate, which shows that DeepSeek generates more percentage of faster codes (about 10\% in average). 

\subsubsection{Retrieval Algorithm}
\begin{table}[thb]\centering
    \caption{Pass@k and speedups achieved by LOOPRAG using loop-aware algorithm compared to BM25 and Weighted Score. "Weighted Score" refers to LOOPRAG with only weighted score (i.e., LAScore w/o BM25).} %
    \resizebox{0.48\textwidth}{!}{
    \fontsize{16}{20}\selectfont
    \begin{tabular}{*{2}{c}*{2}{c}*{2}{c}*{2}{c}}
        \toprule[1.5pt]
        \multirow{2.5}{*}{Method} & \multirow{2.5}{*}{LLMs} & \multicolumn{2}{c}{PolyBench} & \multicolumn{2}{c}{TSVC} & \multicolumn{2}{c}{LORE} \\
        \cmidrule{3-8}
         & & Pass@k  & Speedup  & Pass@k  & Speedup  & Pass@k  & Speedup  \\
        \hline
        \multirow{2}{*}{Loop-aware} & DeepSeek & 70.00 & \textbf{23.97} & \textbf{94.05} & 32.66 & 85.71 & 20.44 \\ 
        & GPT-4 & 70.00 & 14.58 & 84.52 & 31.35 & 87.76 & 14.86 \\
        \midrule
        \multirow{2}{*}{BM25} & DeepSeek & \textbf{73.33} & 16.50 & 88.10 & 28.67 & 89.8 & 11.14 \\
		& GPT-4 & 63.33 & 9.35 & 80.95 & \textbf{42.67} & 87.76 & 9.65 \\
		 \midrule
        \multirow{2}{*}{Weighted Score} & DeepSeek & 70.0 & 18.75 & \textbf{94.05} & 19.61 & \textbf{91.84} & \textbf{21.79} \\
		& GPT-4 & 60.00 & 13.66 & 85.71 & 26.54 & 77.55 & 14.88 \\
        \bottomrule[1.5pt]
    \end{tabular}
    }
    \label{tab:6}
\end{table}

To evaluate the effectiveness of the loop-aware algorithm in retrieval, we compare it with the BM25 algorithm and loop-aware without BM25 score (Weighted Score) to investigate the capability of our method in enhancing code optimization.

As shown in Table \ref{tab:6}, three methods demonstrate similar pass@k. However, the loop-aware algorithm exhibits higher speedup on different aspects corresponding to how it is designed. For example, loop-aware algorithm achieves average speedup improvements of 1.51$\times$ and 1.69$\times$ for BM25 on PolyBench and LORE, respectively. Compared to Weighted Score method, the loop-aware algorithm is higher on PolyBench and TSVC, achieving average speedup improvements of 1.17$\times$ and 1.42$\times$ on PolyBench and TSVC, respectively. 
These results indicate that, by leveraging loop features, the loop-aware algorithm can strike a balance between similarity and diversity, although it achieves lower speedups than BM25 on TSVC and Weighted Score on LORE.

\begin{figure}[thb]
    \centering
    \includegraphics[width=0.48\textwidth]{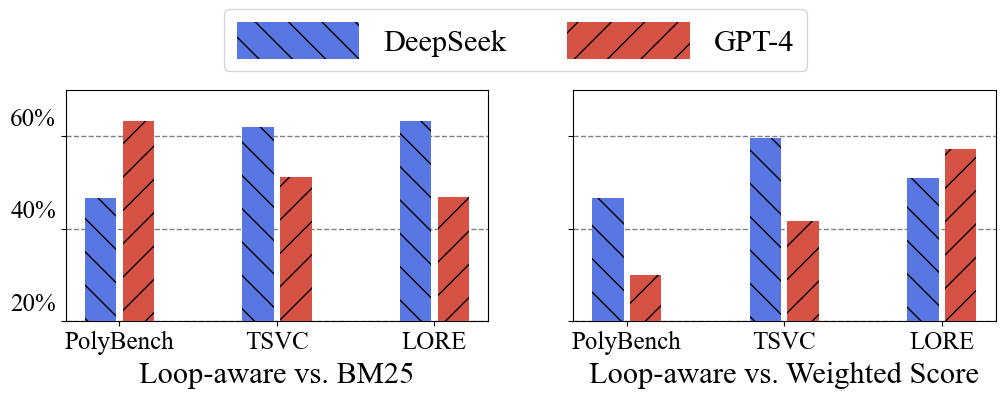}
    \caption{Percentage of faster codes generated by LOOPRAG using loop-aware algorithm compared to BM25 and Weighted Score.}
    \label{fig:9}
\end{figure}

Figure \ref{fig:9} illustrates the percentage of faster codes generated by LOOPRAG using loop-aware algorithm compared to BM25 and Weighted Score. On average, the loop-aware algorithm produces about 50\% more optimized codes with higher speedups than BM25 and Weighted Score. These results highlight its substantial impact on improving code optimization performance. It provides more informative demonstrations than individual scores that focus only on a single aspect. Therefore, we adopt LAScore for its balanced performance and strong capability to handle complex scenarios, effectively guiding LLMs to generate more efficient codes.

\subsubsection{Feedback-based Iterative Generation}

To investigate the impact of feedback mechanism and iterative generation, we measure the \textbf{improvement} in pass@k using the first and second rounds of compilation results feedback, as well as the improvement of both pass@k and speedup using testing results and performance rankings feedback. The first round of compilation results feedback corresponds to the generation in step \ref{enum:1.2}, which is compared with step \ref{enum:1.1}, while the second round refers to step \ref{enum:1.4}, which is compared with step \ref{enum:1.3}. The testing results and performance rankings feedback refer to step \ref{enum:1.4}, which is compared with step \ref{enum:1.2}.

\begin{table}[thb]
    \centering
    \caption{Pass@k \textbf{improvements} after first and second rounds of compilation, and testing results and performance rankings feedback achieved by LOOPRAG.}
    \resizebox{0.48\textwidth}{!}{
        \fontsize{16}{20}\selectfont
        \begin{tabular}{*{2}{c}ccc}
            \toprule[1.5pt]
            \multirow{2}{*}{Feedback} & \multirow{2}{*}{LLMs} & PolyBench & TSVC & LORE \\
            \cmidrule{3-5}
            & & \multicolumn{3}{c}{Pass@k $\uparrow$} \\
            \midrule
            \multirow{2}{*}{First round of compilation} & DeepSeek & 14.44 & \textbf{9.13} & \textbf{20.07} \\ 
            & GPT-4 & \textbf{21.67} & 1.59 & 12.24 \\
            \midrule
            \multirow{2}{*}{Second round of compilation} & DeepSeek & \textbf{3.33} & 1.20 & 4.08 \\ 
            & GPT-4 & \textbf{3.33} & \textbf{4.76} & \textbf{6.12} \\
            \midrule
            Testing results and & DeepSeek & \textbf{5.00} & \textbf{6.24} & \textbf{5.34} \\ 
            Performance rankings & GPT-4 & 1.67 & 4.14 & 5.10 \\
            \bottomrule[1.5pt]
        \end{tabular}
    }
    \label{tab:6.2.3}
\end{table}

Table \ref{tab:6.2.3} illustrates the improvement of pass@k after each iteration with feedback in LOOPRAG. Compilation result feedback successfully increases the average pass@k by 21.39\% (i.e., $(14.44 + 3.33 + 21.67 + 3.33)/2$), 8.34\% and 21.26\% on PolyBench, TSVC and LORE, respectively. Meanwhile, the testing results and performance rankings feedback shows a moderate effect, improving average pass@k by 3.34\%, 5.19\% and 5.22\% on PolyBench, TSVC and LORE, respectively. These results demonstrate the effectiveness of the three types of feedback in guiding LLMs to generate correct codes.

\begin{figure}[thb]
    \centering
    \includegraphics[width=0.36\textwidth]{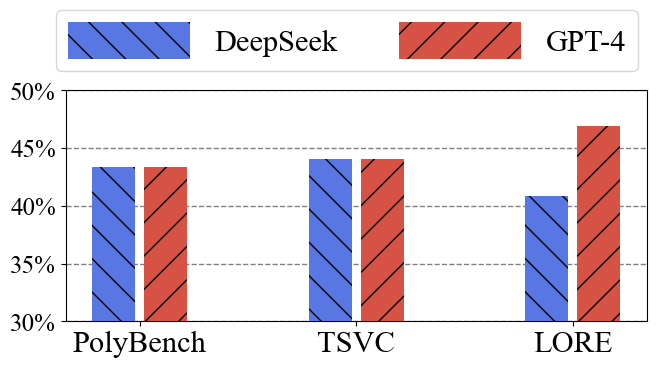}
    \caption{Percentage of faster codes generated by LOOPRAG using testing results and performance rankings feedback.}
    \label{fig:6.2.3}
\end{figure}

As shown in Figure \ref{fig:6.2.3}, the testing results and performance rankings feedback results in an average of 43.33\%, 44.05\% and 43.88\% faster codes for PolyBench, TSVC and LORE, respectively. This contributes to major of the total improvement in the percentage of faster codes shown in Figure \ref{fig:6.1.1}, which demonstrates the effectiveness of these two feedback mechanisms in enhancing the ability of LLMs. 

Overall, these results reveal the impact of our feedback-based iterative generation method in guiding LLMs to generate correct and efficient codes.

\subsection{How does LOOPRAG help improve the quality of loop optimization?} \label{6.5}

To explain why the demonstrations in LOOPRAG exhibit an informative composition of transformation, and how LOOPRAG achieves a balance between similarity and diversity in its demonstrations, we present two example codes \emph{example\_1} and \emph{examples\_2} in Listing \ref{lst:2} and \ref{lst:4} (see optimized version in Appendix \ref{10.6}). These two example codes are retrieved by loop-aware algorithm and used in demonstrations to guide LLMs to generate optimized code \emph{syrk} in PolyBench, as shown in §\ref{2.2}, achieving a 71.60$\times$ speedup compared to the base LLM, GPT-4. 

\begin{figure}[thb]
  \centering
  \begin{minipage}{0.48\textwidth}
\begin{lstlisting}[language=C, caption=Example code \emph{example\_1}., label=lst:2, captionpos=t]
for i = 2 to N
  for j = 0 to M-1
S1: A[i-1][i] = A[i-2][i] + C[i][j] * 6;
  for k = 0 to M-1 
S2: A[k+1][k] = A[i][k] - C[k+1][i] * 4;
\end{lstlisting}
  \end{minipage}
  \hfill
  \begin{minipage}{0.48\textwidth}
\begin{lstlisting}[language=C, caption=Example code \emph{examples\_2}., label=lst:4, captionpos=t]
for i = 0 to L
  for j = 0 to i
S1: A[i][j] = A[i][j] + 6;
  for k = 0 to L 
S2: A[i][k] = - A[k][i] + C[k] - 2;
\end{lstlisting}
  \end{minipage}
\end{figure}

% Both these two codes examples have two statements $S1$ and $S2$, and they use the same arrays $A$ and $C$ in SCoP, resulting in the relation of WAW, WAR and RAW dependence to code.

Comparing these two example codes with target code \emph{syrk}, there are many differences in loop properties, leading to diverse compositions of loop transformations in the optimized versions. For example, although all three codes use the identical arrays $A$, which exhibits the same types of dependence (WAW, WAR and RAW), the indexes of array $A$ differ, causing the dependence to occur in varying patterns.  
In Listing \ref{lst:2}, the indexes of two arrays $A$ in statement $S1$ differ in the first dimension, creating a loop-carried dependence between $S1$ itself across iterations of iterator $i$. However, two arrays $A$ in code \emph{syrk} share the identical indexes (Figure \ref{fig:2}). This subtle difference ultimately results in loop skewing in the optimized version of example code \emph{example\_1}. 
A similar situation occurs in Listing \ref{lst:4}, leading to loop shifting in the optimized version of example code \emph{examples\_2}. 
These differences introduce diversity during retrieval, which helps mitigate the risk of potential misleading guidance for LLMs.

Although differences among example codes lead to various types of loop transformations in demonstrations, loop tiling, loop fusion, and loop interchange are presented in the optimized versions of both example codes.
For example, statements $S1$ and $S2$ in these two example codes are located in the same outer loop nest but reside in different inner loop nest with identical loop bounds. This structure enables the opportunity for loop fusion. Additionally, indexes of arrays $A$ and $C$ in code \emph{example\_1} and array $A$ in code \emph{examples\_2} are not aligned with the order of iterators in loop nests, which is suitable for loop interchange. 
% These transformations contribute to the optimized code \emph{syrk} in Listing \ref{lst:1}.
In this case, these similar characteristics, rooted in loop properties, lead to above profitable loop transformations, which are learnt by LLMs and applied in the optimized code \emph{syrk} in Listing \ref{lst:1}.

In general, these example codes retrieved by LOOPRAG successfully exhibit both similarity and diversity in loop properties. They result in demonstrations in LOOPRAG, which contain an appropriate composition of loop transformations in the optimized versions of example codes, ultimately improving the quality of loop optimization.

\section{Related Work}

\textbf{Loop Dataset Construction.}
A large-scale dataset serves as a valuable resource for both training code optimization models and evaluating compilers. Several recent approaches have been proposed to construct loop datasets. 
LORE \cite{8167779} maintains a large amount of C language \emph{for} loop nests extracted from benchmark suites, libraries, and real-world applications used for evaluating compilers.
ANGHABENCH \cite{9370322} employs a combination of web crawling and type inference to generate C programs from open-source repositories for code size reduction. 
YARPGen \cite{10.1145/3591295} is a generative fuzzer that presents a three-step loop generation policy to trigger loop optimizations for finding compiler bugs. 
Baghdadi et al. \cite{MLSYS2021_d9387b6d} develop a random code generator to create synthetic data for training a cost model for code optimization, while Mezdour et al. \cite{10.1145/3578360.3580257} follow the same methodology to produce a dataset for train a deep learning model for loop interchange. 
COLA-Gen \cite{berezov_et_al:OASIcs.PARMA-DITAM.2022.3} introduces a loop generator capable of creating synthetic training datasets in a parametric way for code optimization.
% In contrast, LOOPRAG synthesizes example codes considering loop properties for loop structure, dependence and array access, which are applicable for various composition of loop transformations and lead to high efficiency improvement, which tackles the challenge of code generation with complexity in code syntax and semantics.
However, current loop generation methods offer limited diversity in loop properties and do not effectively trigger loop optimizations. Therefore, we introduce a parameter-driven approach to address this issue.

\textbf{LLM-based Code Optimization.}
Due to deep semantic understanding of code, LLMs have demonstrated significant potential in various code optimization tasks \cite{gong2025languagemodelscodeoptimization}.
Cummins et al. \cite{cummins2023largelanguagemodelscompiler, cummins2024metalargelanguagemodel, grubisic2024compilergeneratedfeedbacklarge} aim to optimize the code size of LLVM assembly by using compiler-generated feedback to guide the LLM.
RALAD \cite{10691855} proposes a retrieval-augmented LLM-based approach to optimize high-level synthesis (HLS) programs. 
SBLLM \cite{gao2024search} enables LLMs to iteratively refine and optimize code through evolutionary search. 
PCAOT \cite{rosas2024aioptimizecodecomparative} utilizes instruction prompting and chain-of-thought to direct LLMs in improving code performance using OpenMP, while LLM-Vectorizer \cite{taneja2024llmvectorizerllmbasedverifiedloop} employs an AI agent-based approach with test-driven feedback to harness LLMs for generating vectorized code.
However, due to the lack of performance feedback or intrinsic mechanisms—such as cost models or heuristics found in compilers—to guide effective strategies for code optimization, LLMs often suggest suboptimal or incorrect optimizations. Our work aims to bridge this gap by providing informative demonstrations and feedback mechanisms, iteratively guiding LLMs to generate correct and efficient code.
% \vspace{0.5cm} % 调整第一章后的间距

% 第二章前后调整间距
% \vspace{-0.75cm} % 调整第二章前的间距
\section{Conclusion}
In this paper, we propose LOOPRAG, a framework designed to enhance loop transformation optimization on SCoP through retrieval-augmented generation. Our study emphasizes the critical role of providing informative demonstrations to guide LLMs in performing code optimization. To address this challenge, we introduce a parameter-driven method to synthesize a dataset as the demonstration source. To effectively obtain valuable demonstrations, we propose a loop-aware algorithm for code retrieval. Additionally, we present a feedback-based iterative generation method to guide LLMs in generating correct and efficient code. Evaluation results demonstrate that LOOPRAG outperforms the state-of-the-art compilers, representative LLMs and LLM-based models, showcasing significant improvements in code optimization. For future work, we plan to incorporate more code optimization techniques and refine the generation steps, which could overcome the current limitation on SCoP, improve the soundness of the optimized code and make the whole procedure automatic.

% \section{Acknowledgments}

\bibliographystyle{ACM-Reference-Format}
\bibliography{main}

\newpage

\appendix

\section{Loop Parameters in Dataset Synthesis} \label{10.1}

In this appendix, we explain the meanings and ranges of the ten loop parameters used to configure loop properties in our parameter-driven method:

\begin{enumerate}[1)]
\item \textbf{Iterator Bounds} specifies the probability of iterators being present in loop bounds, which decreases by half at each subsequent level in the loop nest. The value of this parameter is randomly selected from \{20\%, 40\%, 60\%\}.
\item \textbf{Loop Depth} specifies the maximum loop depth of SCoP. The value of this parameter is randomly selected from 2 to 4.
\item \textbf{Statement Index} specifies the maximum number of loop branches in every level of loop nests, which determines the density of statements in loops. The value of this parameter is randomly selected from 1 to 3.
\item \textbf{Number of Statements} specifies the number of statements in SCoP. The value of this parameter is randomly selected from 1 to 6.
\item \textbf{Dep Distance} specifies the maximum absolute value of the bounds in each dimension of the dependence distance vector. The value of this parameter is randomly selected from 1 to 2.
\item \textbf{Read Dep} specifies the maximum number of WAR and RAW dependence per statement. The value of this parameter is randomly selected from 1 to 3.
\item \textbf{Write Dep} specifies the probability that WAW dependence exists for each statement. The value of this parameter is randomly selected from \{20\%, 40\%, 60\%\}.
\item \textbf{Array List} specifies the number of alternative arrays for both write and reads in each statement, which indicates the variety of arrays. The value of this parameter is randomly selected from 1 to 3.
\item \textbf{Read Array} specifies the maximum number of reads to array per statement. The value of this parameter is randomly selected from \{1, 3, 5\}
\item \textbf{Array Indexes} specifies the maximum absolute value of the bounds of constant coefficient for array indexes matrix. The value of this parameter is randomly selected from 1 to 2.
\end{enumerate}

Although accessing the same indexes of an array across different iterations can result in dependence, such situations occur rarely when the values of array-related parameters (i.e., \textbf{Array List}, \textbf{Read Array} and \textbf{Array Indexes}) are assigned randomly. Hence, we utilize dependence-related parameters (i.e., \textbf{Dep Distance}, \textbf{Read Dep} and \textbf{Write Dep}) to manually configure dependence, beyond those naturally created by array accesses, for each statement, thereby diversifying example codes.
Besides, we constrain the dependence distance to be constant, and generate array indexes whose iterator coefficients are set to 1.

\section{Example Code Synthesis Algorithm in Dataset Synthesis} \label{10.2}

\begin{algorithm}[!tb]
\caption{Example Code Synthesis from Loop Parameters}
\label{algo:1}
\small
\SetKw{in}{in}
\KwIn{\textit{IteratorBound, LoopDepth, StmtIndex, NStmts, DepDistance, ReadDep, WriteDep, ArrayList, ReadArr, ArrIndexes, NameList, SizeList}}
\KwOut{\textit{Code}} 
% \textit{ArrsNoDep, ArrsDep, Iterators, Bounds, StmtsExpr, LoopsExpr, ScheduleTree} = [], [], [], [], [], []\;

\textit{SchedulesMtx} = \textbf{Rand}(\textit{NStmts, LoopDepth, StmtIndex})\; \label{algo:1.1}

\textit{SchedulesMtx} = \textbf{ReOrder}(\textit{SchedulesMtx})\; \label{algo:1.2}

\textit{ItBounds} = \textbf{GenItBounds}(\textit{IteratorBound}, \textit{SchedulesMtx})\; \label{algo:1.3}

\For {i = $0$ \KwTo \textit{NStmts} - $1$}{ \label{algo:1.4}
    \textit{Names} = \textit{NameList}[ArrayList]\; \label{algo:1.5}
    \textit{Indexes} = \textbf{GenIndex}(ArrIndexes)\; \label{algo:1.6}
    \textit{ArrsNoDep}[i].\textit{append}([\textit{Names}, \textit{Indexes}])\; \label{algo:1.7}

    \textit{SourceIds} = \textbf{DepAssign}(WriteDep,ReadDep,NStmts)\; \label{algo:1.8}
    \textit{ArrsDep}[i].\textit{append}([\textit{SourceIds}])\; \label{algo:1.9}
}

\textit{Arrs} = \textbf{PriorityAssign}(\textit{ArrsNoDep}, \textit{ArrsDep})\; \label{algo:1.10}

\For {i = $0$ \KwTo \textit{NStmts} - $1$}{ \label{algo:1.11}
    \For {\textit{Arr} \in \textit{Arrs}[i]}{ \label{algo:1.12}
        \textit{ID} = \textbf{SearchSource}(\textit{Arr}[\textit{SourceId}])\; \label{algo:1.13}
        \textit{ArrsSize}[\textit{Arr}[\textit{Name}]] = \textbf{SizeAssign}(\textit{Arr}, \textit{SizeList})\; \label{algo:1.14}
    }
}

\For {i = $0$ \KwTo \textit{NStmts} - $1$}{ \label{algo:1.15}
    \textit{Iterators.append}(\textbf{ExtractIterator}(\textit{SchedulesMtx}[\textit{i}]))\; \label{algo:1.16}
    \textit{ScheduleTree.append}(\textit{SchedulesMtx}[\textit{i}])\; \label{algo:1.17}
}

\textit{Arrs} = \textbf{CheckArrDep}(\textit{Arrs})\; \label{algo:1.18}
\textit{Bounds} = \textbf{CreateBoundsData}(\textit{ItBounds})\; \label{algo:1.19}

\For {i = $0$ \KwTo \textit{NStmts} - $1$}{ \label{algo:1.20}
    \textit{Arrs}[i][\textit{Indexes}] = \textbf{ReGenerateIndex}(DepDistance)\; \label{algo:1.21}
    \textit{Bounds.append}(\textbf{GetBounds}(\textit{Arrs}[i]))\; \label{algo:1.22}
    \textit{StmtsExpr.append}(\textbf{ConvertToExprs}(\textit{Arrs}[i]))\; \label{algo:1.23}
}

\For {\textit{ArrSize} \textbf{in} \textit{ArrsSize}}{ \label{algo:1.24}
    \textit{Bounds.append}(\textbf{GetBounds}(ArrSize))\; \label{algo:1.25}
}

\For {It \in Iterators}{ \label{algo:1.26}
    \textit{LB, UB} = \textbf{CalcBound}(\textit{It, Bounds})\; \label{algo:1.27}
    \textit{LoopsExpr.append}(\textbf{GenForLoop}(\textit{LB, UB}))\;  \label{algo:1.28}
}

\textit{Structure} = \textbf{UpdateTree}(\textit{ScheduleTree})\; \label{algo:1.29}
\textit{SCoP} = \textbf{GetLoop}(\textit{Structure, LoopsExpr, StmtsExpr})\; \label{algo:1.30}
\textit{Code} = \textbf{GetCode}(\textit{SCoP})\; \label{algo:1.31}

\Return \textit{Code} \label{algo:1.32}
\end{algorithm}

Algorithm \ref{algo:1} shows the procedure of generating example codes from loop parameters in LOOPRAG. It starts with ten parameters and two lists containing name and size prepared for arrays.
Initially, loop schedule are configured with random value and dimension for each statement, and are then reordered considering the partial order (line \ref{algo:1.1}-\ref{algo:1.2}). The construction of schedule allocates the order of iterators in loop nests, allowing the generation of iterator bound (line \ref{algo:1.3}).

Arrays named $ArrsNoDep$, which act as the alternatives for arrays named $ArrsDep$, are generated with array name and indexes (line \ref{algo:1.5}-\ref{algo:1.7}). Arrays named $ArrsDep$ are created based on dependence-related parameters (lines \ref{algo:1.8}-\ref{algo:1.9}). If the array is related to a WAW dependence, $ArrsDep$ is generated as the target array with the ID of statement which the source array in this WAW dependence locates. 
In the case of WAR or RAW dependence, $ArrsDep$ pertains to the read to array in WAR or RAW dependence, and contains the ID of statement corresponding to the write to array. 

All arrays are assigned based on priority to eliminate $ArrsNoDep$ alternatives if necessary (line \ref{algo:1.10}). 
The remaining arrays build the relationship through the statement ID from $ArrsDep$ based on dependence, and randomly obtain their size from the list (line \ref{algo:1.12}-\ref{algo:1.14}).
These arrays will be checked for the validity of related dependence, and resolve the contradictions such as circular dependence (line \ref{algo:1.18}). The indexes of arrays related to valid dependence are then regenerated (line \ref{algo:1.21}). 

The value from array size, array indexes, and loop iterators (line \ref{algo:1.25}, line \ref{algo:1.22} and line \ref{algo:1.19}, respectively) are used to calculate the loop bounds (line \ref{algo:1.27}). Once there is no solution, the generation procedure will break off.

All texture information, such as iterator and array name, is collected for statement expressions (line \ref{algo:1.23}) and loops expressions (line \ref{algo:1.28}). The SCoP of the code is constructed from schedule tree (line \ref{algo:1.29}) and finalized by adding initialization, output extraction, and the main function in PolyBench style, resulting in a complete program (line \ref{algo:1.30}-\ref{algo:1.31}).
Besides, each file is accompanied with header files containing definitions of other functional modules, such as macros defined for data types and the values of global parameters.

\section{SCoP Extraction} \label{10.8}

The SCoP detection in LOOPRAG and PLuTo is implemented using Clan. Clan identifies the directives \emph{\#pragma scop} and \emph{\#pragma endscop} in the program and checks whether the enclosed code region is syntactically valid. 

As for Graphite and Polly, which also treat SCoPs as regions for optimization, their methods of SCoP detection differ from Clan. Unlike Clan's primarily syntactic checks, both Graphite and Polly enforce stricter rules, analyzing whether a loop region fully satisfies the requirements of a valid SCoP. In particular, all operations within a SCoP (including function calls) must be side-effect free---that is, restricted to expressions that do not modify any state outside the SCoP.

The codes in PolyBench and LORE follow the standard SCoP design, whereas each TSVC kernel contains a \emph{dummy} function call in the outer loop. In such cases, SCoP detection fails in Graphite and Polly, unless the \emph{dummy} function is explicitly ensured to be side-effect free. 

Unlike Polly's precise SCoP detection, Clan lacks rigorous verification rules, such as pointer aliasing checks or function side-effect analysis. Instead, it inspects only the raw code within the annotated SCoP region, implicitly assuming that all functions enclosed by \emph{\#pragma scop} and \emph{\#pragma endscop} are side-effect free. Consequently, Clan omits explicit function analysis (e.g., the \emph{dummy} function call in the outer loop in TSVC kernel) and simply treats such functions as statements within the SCoP.

We annotate the \emph{dummy} function with the \emph{\_\_attribute\_\_((pure))} qualifier to instruct Polly to treat the function as free of side effects and to ignore its return value during analysis. This annotation ensures that calls to \emph{dummy} are considered side-effect free, thereby enabling the successful identification of the surrounding loop structure as a valid SCoP. Without this qualifier, the function call would be regarded as a potential source of side effects, causing both Graphite and Polly to discard the SCoP candidate.

However, this qualifier is ineffective for Graphite: applying  \emph{\_\_attribute\_\_((pure))} causes the outer computational loop to be eliminated by Dead Code Elimination (DCE). Consequently, we exclude Graphite from the TSVC comparison to ensure fairness.

\section{Loop Feature Extraction} \label{10.3}

\begin{figure}[thb]
    \centering
    \includegraphics[width=\linewidth]{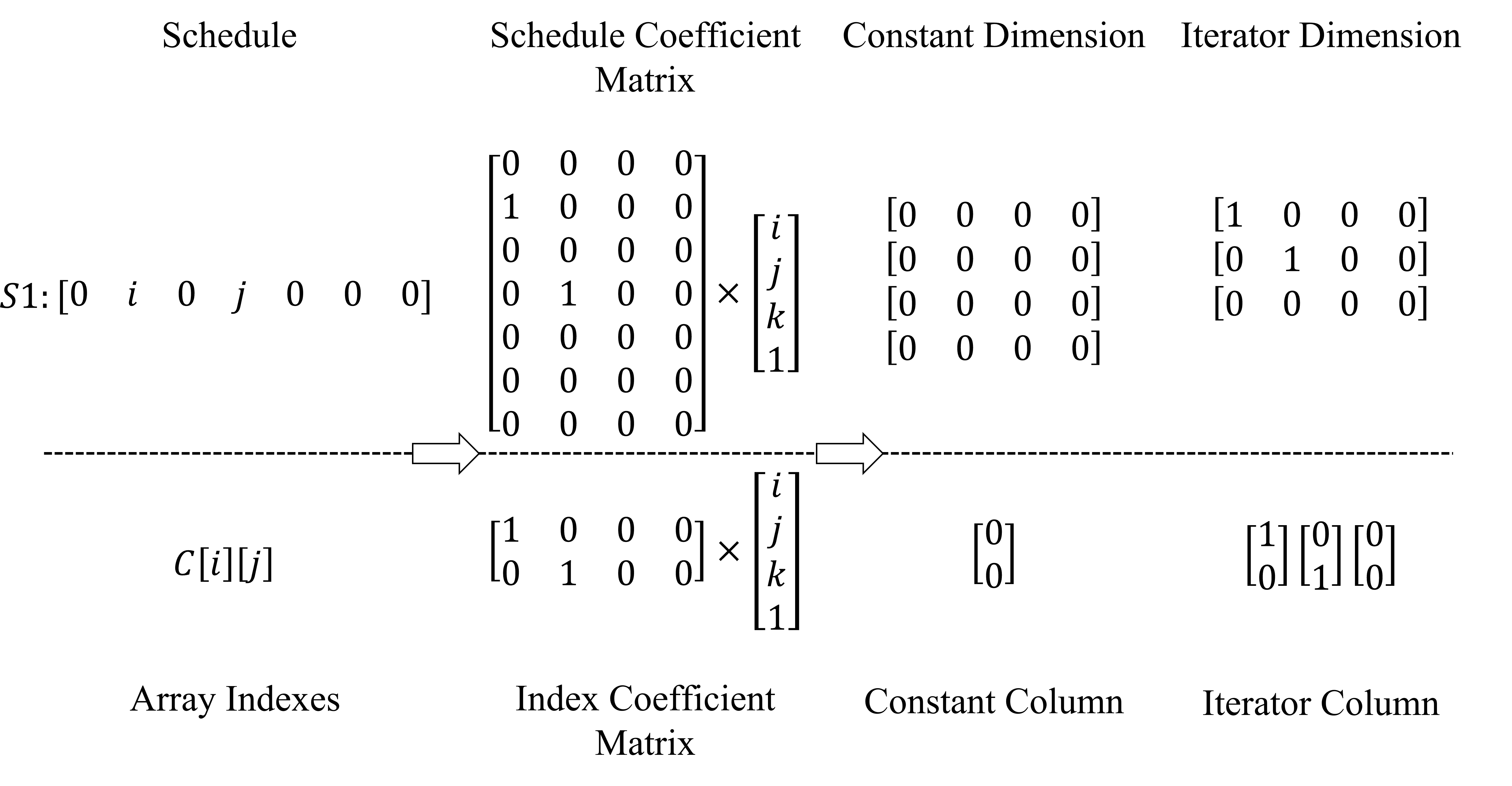}
    \caption{Example of feature extraction procedure for original code \emph{syrk} in Figure \ref{fig:2}.}
    \label{fig:4.2}
\end{figure}

LOOPRAG proposes a loop-aware algorithm for retrieval considering loop properties. We select and abstract loop schedule and array indexes properties as loop features. In these properties, loop schedule is extracted in a 2d+1 matrix form, and array indexes are presented as matrix containing coefficient of iterators and constants in each dimension (Figure \ref{fig:4.2}).    

% constant and iterator dimension of indexes matrix of reads and writes to array which is related to dependence

% iterator dimension of indexes matrix of reads and writes to array which is not related to dependence.

Loop schedule is separated into constant and iterator dimensions to present the partial order of this statement at each level of the loop nests and the order of iterators in its surrounding loop nest. 
Array indexes are differentiated for reads and writes to array to highlight the difference of their significance. Each indexes are split into constant and iterator columns, and then selected according to dependence. Loop-independent dependence relates only to constant columns in array indexes, while loop-carried dependence considers both.
Besides, all meaningless zero columns in the indexes matrix are removed to facilitate matching between array indexes with different number of dimensions. 

\section{Prompts} \label{10.4}

We guide LLMs to apply loop transformation optimization, learn from demonstrations and utilize the feedback information through the following prompts.

\subsection{Base Prompt}

\begin{adjustwidth}{1pt}{1pt}
\begin{quote} \label{q:1}
\itshape
As a compiler, given the C program below, improve its performance using meaning-preserving loop transformation methods:

\{target\_code\}

Here are some generation rules:
1. Provide one optimized code.
2. Do not include the original C program in your response.
3. Do not define new function. 
4. Existed variables do not need to be redefined. If you generate new variable for computing, please use the double type. 
5. Put your code in markdown code block.
\end{quote}
\end{adjustwidth}

This prompt is used for the baseline LLMs in §\ref{6.2.2} to generate optimized code. $\{target\_code\}$ refers to target code. The rules in the last few lines are also added in all prompts later, which are used to constrain the output of generation. They can largely reduce runtime error in generation and ensure the completion from generated SCoP to a whole program for testing. The double type is used the same as original code settings in benchmark.

\subsection{Prompt with Demonstration}

\begin{adjustwidth}{1pt}{1pt}
\begin{quote} \label{q:2}
\itshape
// original code

\{ori\_code\}

// optimized code

\{opt\_code\}

Please analyze what meaning-preserving loop transformation methods are used in above examples, and tell me what you learn.

please use appropriate methods you learn from examples to improve its performance:

\{target\_code\}
\end{quote}
\end{adjustwidth}

This prompt is used in generation step 1 as the first attempt to generate optimized code. $\{ori\_code\}$ and $\{opt\_code\}$ refer to example code and optimized version of example code. We specifically add the words "$analyze$" and "$learn$" to enhance the optimization ability of LLMs.

\begin{figure*}[thb]
    \centering
    \includegraphics[width=0.96\textwidth]{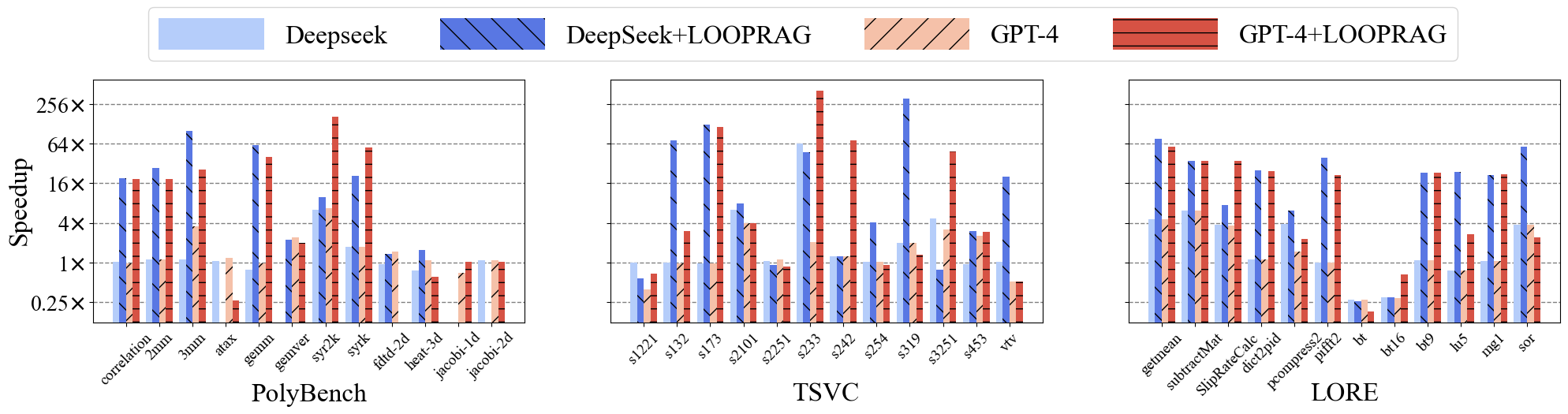}
    \caption{Speedups achieved by LOOPRAG and base LLMs.}
    \label{fig:4}
\end{figure*}

\subsection{Compilation Results Feedback Prompt}

\begin{adjustwidth}{1pt}{1pt}
\begin{quote} \label{q:3}
\itshape
This optimized version:

\{last\_code\}

did a wrong transformation from the source code, resulting in a compilation error. 
This is the compiler error message:

\{error\}

Please check the optimized code and regenerate it.
\end{quote}
\end{adjustwidth}

This prompt is used in generation step 2 and 4 for the first and second rounds of compilation results feedback, accompanied with the messages in previous chat. $\{last\_code\}$ and $\{error\}$ indicate the optimized codes generated in the last step and the error information provided by compiler during compilation.

\subsection{Testing Results and Performance Rankings Feedback Prompt}

\begin{adjustwidth}{1pt}{1pt}
\begin{quote} \label{q:4}
\itshape
Available Example [\{idx\}]:

\{feedback\_code\}

Failed Example [\{idx\}]:

\{feedback\_code\}

...

The above examples are optimized by LLMs using meaning-preserving loop transformation methods. 
Available examples pass compilation, execution and equivalence checks; failed examples do not. 
Here is the original code:

\{target\_code\}

Performance rank result (">" means better than):

\{idx\} > ...

Failed: \{idx\}, ...

Task:
Analyze why available examples succeeded and failed examples broke correctness. Improve the performance of original code using the highest-impact meaning-preserving loop transformation methods learnt from the ranked examples.
\end{quote}
\end{adjustwidth}

Step 3 in generation utilizes both testing results and performance rankings as feedback in the above prompt with the messages in previous chat. Only the codes that pass testing will be included this prompt, along with the ranking results to provide feedback simultaneously. $\{idx\}, \{feedback\_code\}$, and $\{target\_code\}$ denote the ID of previous optimized codes, the content of previous optimized codes, and the content of original target code, respectively. 
In the rankings, "..." represents multiple IDs of previously optimized codes and the original target code. If all previous optimized codes fail, the prompt will revert to Prompt \ref{q:1}.
% \vspace{0.5cm} % 调整第一章后的间距

% 第二章前后调整间距
% \vspace{-0.75cm} % 调整第二章前的间距
\section{Benchmark Speedup Analysis of LOOPRAG Compared to Base LLMs} \label{10.5}

To further analyze and compare the performance improvements achieved by LOOPRAG and base LLMs, we present a subset of benchmarks that demonstrate the speedup results compared to GCC in Figure \ref{fig:4}. In many cases, LOOPRAG achieves speedup of up to over 256$\times$, significantly surpassing those of base LLMs, which do not exceed 8$\times$.

By investigating the difference between benchmark suites, we observe that there are many codes in TSVC which have lower speedups compared to codes in PolyBench and LORE. However, the average speedup metric for TSVC in Table \ref{tab:1}, compared to GCC, are higher than those of the other benchmark suites.  
It suggests that several optimized codes generated by LOOPRAG in TSVC reach extremely high speedups, distorting the metric of average speedup.
For example, code $s233$ and $s319$ reach 412.22$\times$ and 307.62$\times$ for LOOPRAG using DeepSeek and GPT-4. These optimized codes utilize loop tiling and loop interchange to enhance data locality and apply parallelism through OpenMP directive, resulting in extremely high average speedup in TSVC. 

However, the average speedup improvement achieved by LOOPRAG compared to base LLMs for TSVC is lower than that for PolyBench and LORE, as mentioned in §\ref{6.2.2}. The percentage of faster codes for TSVC in Figure \ref{fig:6.1.1} is also not significantly prominent. 
It means that the codes in TSVC is relatively easier for LLMs to analyze and optimize, which makes it harder for LOOPRAG to outperform base LLMs. Consequently, the speedup improvement for LOOPRAG compared to LLM-based methods primarily attribute to codes with complex loop properties, which better reflect the value of loop transformations. 

\section{Case Studies}  \label{10.6}

\subsection{Case Study for Optimized Versions}
\begin{figure}[thb]
  \centering
  \begin{minipage}{0.48\textwidth}
\begin{lstlisting}[language=C, caption=Optimized Version \emph{example\_1}., label=lst:10.6.1, captionpos=t]
#pragma omp parallel
for t1 = 0 to N/32
 for t2 = 0 to (t1+N)/32
  for t3 = 32*t1 to 32*t1+32
   for t4 = 32*t2 to 32*t2+32
S1: A[t3-1][t3] = A[t3-2][t3] + C[t3][t4-t3] * 6;
    if t3 < M & t4 < N
S2:  A[t3+1][t3] = A[t4][t3] - C[t3+1][t4] * 4;
\end{lstlisting}
  \end{minipage}
  \hfill
  \begin{minipage}{0.48\textwidth}
\begin{lstlisting}[language=C, caption=Optimized Version \emph{examples\_2}., label=lst:10.6.2, captionpos=t]
#pragma omp parallel
for t1 = 0 to 2*L/32
 for t2 = max(0, t1-L/32) to min(L/32, t1)
  for t3 = 32*t1 to 32*t1+32
   for t4 = 32*t2 to 32*t1+32
    if t3-t4 < t4
S1:  A[t4][t3-t4] = A[t4][t3-t4] + 6;
S2: A[t4][t3-t4] = - A[t3-t4][t4] + C[t3-t4] - 2;
\end{lstlisting}
  \end{minipage}
\end{figure}

% 要讲目的
% 要独立可懂，直接重新解释一下
To illustrate the loop transformations contained in the demonstrations, we present the optimized versions of example code \emph{examples\_1} and \emph{examples\_2} in Listing \ref{lst:10.6.1} and \ref{lst:10.6.2}.
The transformations include loop tiling, loop fusion, loop interchange, loop skewing and loop shifting as mentioned in §\ref{6.5}.

Loop tiling divides and separates iterators to create characteristics like $/32$ and $t1,t2,t3,t4$. 
Loop fusion rearranges statements so that they share the same loop nest.
Loop interchange modifies the order of surrounding loop nests. For example, the surrounding loop nests of $S2$ in Listing \ref{lst:10.6.1} are $t1,t2,t3,t4$ with the condition $t3 < M \& t4 < N$, corresponding to the order of "$k,i$", which is exchanged compared to the order of "$i,k$" in Listing \ref{lst:2}.
Loop skewing involves features such as $(t1+N)/32$ in Listing \ref{lst:10.6.1} and $max(0, t1-L/32)$ Listing \ref{lst:10.6.2} to reassign the number of iterations in loop nests. 
Loop shifting is presented as the condition $t3-t4 < t4$ in Listing \ref{lst:10.6.2}, to align the iterations of different statements.

These loop transformations in the optimized versions provide potential composition of loop transformations as demonstrations for code \emph{syrk}. The impact of diversity introduced by loop skewing and loop shifting in these two optimized versions can be overshadowed by the composition of loop tiling, loop fusion and loop interchange, since they are more commonly found in the optimized versions of retrieved example codes. Finally, these demonstrations guide LLMs to generate efficient optimized code \emph{syrk} as shown in Figure \ref{fig:2}.

\subsection{Case Study for How LOOPRAG Outperforms Base LLMs}

\begin{figure}[thb]
  \centering
  \begin{minipage}{0.48\textwidth}
\begin{lstlisting}[language=C, caption=Original code \emph{gemm}., label=lst:6, captionpos=t]
for i = 0 to NI
  for j = 0 to NJ
S1: C[i][j] *= beta;
  for k = 0 to NK
    for j = 0 to NJ   
S2:   C[i][j] += alpha * A[i][k] * B[k][j];
\end{lstlisting}
  \end{minipage}
  \hfill
  \begin{minipage}{0.48\textwidth}
\begin{lstlisting}[language=C, caption=Optimized code \emph{gemm} generated by base LLM., label=lst:7, captionpos=t]
for i = 0 to NI
  for j = 0 to NJ
S1: double temp = C[i][j] * beta;
    for k = 0 to NK
S2:   temp += alpha * A[i][k] * B[k][j];
S3: C[i][j] = temp;
\end{lstlisting}
  \end{minipage}
  \hfill
  \begin{minipage}{0.48\textwidth}
\begin{lstlisting}[language=C, caption=Optimized code \emph{gemm} generated by LOOPRAG., label=lst:8, captionpos=t]
#pragma omp parallel
for t1 = 0 to NI/32
 for t2 = 0 to NJ/32
  for t3 = 32*t1 to 32*t1+32
   for t4 = 32*t2 to 32*t2+32
S1: C[t3][t4] *= beta;
 for t2 = 0 to NK/32
  for t3 = 32*t1 to 32*t1+32
   for t4 = 32*t2 to 32*t2+32
    for j = 0 to NJ
S2:  C[t3][t4] += alpha * A[t3][j] * B[t4][j];
\end{lstlisting}
  \end{minipage}
\end{figure}

To demonstrate how LOOPRAG outperforms base LLMs, we present the original code \emph{gemm} alongside its optimized codes generated by base LLM (DeepSeek) and LOOPRAG in Listing \ref{lst:6}, \ref{lst:7} and \ref{lst:8}. Code \emph{gemm} is nearly identical to code \emph{syrk} (Figure \ref{fig:2}) in text features. The only difference lies in the array indexes of the last read to array in $S2$, while code \emph{syrk} uses array $A[j][k]$ and code \emph{gemm} uses array $B[k][j]$.

This subtle change creates a completely different environment for searching the appropriate composition of loop transformations. In Figure \ref{fig:2}, the iteration traverses iterator $j$ first, followed by $k$ for $A[j][k]$, which prevents continuous access when array $A$ is stored in row-major order as default. However, this issue does not exist in code \emph{gemm}, leading to difference in data locality and making the same composition of loop transformations unsuitable for achieving optimal efficiency.

As shown in Listing \ref{lst:7}, DeepSeek applies loop fusion and loop interchange to code \emph{gemm}, and introduces a new temporary variable $temp$ to reduce the number of accesses to $C[i][j]$. This results in a 0.20$\times$ speedup compared to the original code, as illustrated in Figure \ref{fig:4}. In contrast, LOOPRAG achieves a 60.61$\times$ speedup over original code by leveraging loop tiling and $\#pragma\ omp\ parallel$ directive, learned from the optimized version of example codes. This represents a 303.05$\times$ improvement over DeepSeek. Although DeepSeek’s speedup can increase to 3.86$\times$ after adding $\#pragma\ omp\ parallel$, it remains significantly lower than LOOPRAG’s performance. The base LLM GPT-4 outperforms DeepSeek on this target code but achieves only a 0.76$\times$ speedup, compared to LOOPRAG’s 44.56$\times$, further highlighting LOORAG’s the success. These results provide strong evidence supporting our explanation for Figure \ref{fig:1} and demonstrate the effectiveness of LOOPRAG in guiding LLMs to optimize code.

\section{Discussion}\label{10.7}

% 一个错误实例没有价值，对错误进行总结、分类分析才有意义，可以参考ASE 2023 中，P19右下角错误分析的写法，以及DISCUSSION的写法

Although LOOPRAG outperforms industry compilers, LLM-based methods and PLuTo in code optimization, it remains significant potential for improvement in future work.

LOOPRAG utilizes a parameter-driven method to synthesize example codes. However, existing example codes still lack certain computing characteristics. For examples, Listing \ref{lst:10.6.3} provides the original code \emph{jacobi-2d} from PolyBench, which achieves only a 0.58$\times$ speedup for LOOPRAG.

\begin{figure}[thb]
  \centering
  \begin{minipage}{0.48\textwidth}
\begin{lstlisting}[language=C, caption=Original code \emph{jacobi-2d}., label=lst:10.6.3, captionpos=t]
for t = 0 to T
 for i = 1 to N
  for j = 1 to N
S1:B[i][j] = A[i][j] + A[i][j-1] + A[i][1+j] + A[1+i][j] + A[i-1][j];
 for i = 1 to N
  for j = 1 to N  
S2:A[i][j] = B[i][j] + B[i][j-1] + B[i][1+j] + B[1+i][j] + B[i-1][j];
\end{lstlisting}
  \end{minipage}
\hfill
  \begin{minipage}{0.48\textwidth}
\begin{lstlisting}[language=C, caption=Optimized Code \emph{jacobi-2d}., label=lst:10.6.4, captionpos=t]
for t1 = 0 to T/32
#pragma omp parallel
 for t2 = 0 to N/32
  for t3 = 32*t2 to 32*t2+32
   for t4 = 0 to N/32
S1: B[t3][t4] = A[t3][t4] + A[t3][t4-1] + A[t3][1+t4] + A[1+t3][t4] + A[t3-1][t4];
#pragma omp parallel
 for t2 = 0 to N/32
  for t3 = 32*t2 to 32*t2+32
   for t4 = 0 to N/32
S2: A[t3][t4] = B[t3][t4] + B[t3][t4-1] + B[t3][1+t4] + B[1+t3][t4] + B[t3-1][j];
\end{lstlisting}
  \end{minipage}
\end{figure}

% loop skewing 对structure要求高
% retrieval balance 依然要提高

The lower speedup is attributed to the lack of appropriate loop transformations. In Listing \ref{lst:10.6.3}, the reads to arrays in $S1$ and $S2$ exhibit a specific pattern of array indexes for $[i][j]$, $[i+1][j]$, $[i-1][j]$, $[i][j+1]$ and $[i][j-1]$. This pattern is characteristic of $stencil$ computations, which can be optimized using loop skewing to achieve wavefront parallelism. However, LOOPRAG only applies loop tiling in its optimized code. A similar situation occurs for $stencil$ computations in code \emph{fdtd-2d}, \emph{heat-3d} and \emph{jacobi-1d} from PolyBench, as illustrated in Figure \ref{fig:4}.

This computing characteristic is relatively sophisticated and rare in other two benchmark suites, but is widely used in real-world applications, such as image processing. 
In this case, we need to refine loop property configuration in synthesis, improve loop feature extraction in retrieval, or introduce templates tailored to the above computing characteristics to produce more diverse example codes. Moreover, LOOPRAG lacks sound semantic equivalence guarantees, relying on extensive testing rather than provable correctness. Additionally, the process can be time-consuming, and LLM outputs show significant variance even with the temperature set to 0.

This lack of soundness prevents LOOPRAG from functioning as a fully automatic tool for code optimization. In general, we need to achieve better balances between diversity and legality during dataset synthesis, diversity and similarity during retrieval, and more efficient methods for validating correctness and performance during generation.

% Besides, more code optimization techniques could be added into demonstration to broaden the power of our framework.

\end{document}